# Scattering Interference Signature of a Pair Density Wave State in the Cuprate Pseudogap Phase

Shuqiu Wang[1§,] Peayush Choubey[2,3§], Yi Xue Chong[4§], Weijiong Chen[1], Wangping Ren[1], H. Eisaki[5], S. Uchida[5], P.J. Hirschfeld[6] and J.C. Séamus Davis[1,4,7,8*]

1.  *Clarendon Laboratory, University of Oxford, Oxford, OX1 3PU, UK*
2.  *Institut für Theoretische Physik III, Ruhr-Universität Bochum, D-44801 Bochum, Germany.*
3.  *Department of Physics, Indian Institute of Technology (Indian School of Mines), Dhanbad, Jharkhand-826004, India.*
4.  *LASSP, Department of Physics, Cornell University, Ithaca NY 14850, USA*
5.  *Inst. of Advanced Industrial Science and Tech., Tsukuba, Ibaraki 305-8568, Japan.*
6.  *Department of Physics, University of Florida, Gainesville, FL, USA*
7.  *Department of Physics, University College Cork, Cork T12 R5C, Ireland*
8.  *Max-Planck Institute for Chemical Physics of Solids, D-01187 Dresden, Germany*
§   *These authors contributed equally to this project.*
*   *Correspondence and requests for materials should be addressed to J.C.S.D. (email: jcseamusdavis@gmail.com )*

## Abstract

An unidentified quantum fluid designated the pseudogap (PG) phase is produced by electron-density depletion in the $CuO_2$ antiferromagnetic insulator. Current theories suggest that the PG phase may be a pair density wave (PDW) state characterized by a spatially modulating density of electron pairs. Such a state should exhibit a periodically modulating energy gap $\Delta_P(\boldsymbol{r})$ in real-space, and a characteristic quasiparticle scattering interference (QPI) signature $\Lambda_P(\boldsymbol{q})$ in wavevector space. By studying strongly underdoped $Bi_2Sr_2CaDyCu_2O_8$ at hole-density ~0.08 in the superconductive phase, we detect the $8a_0$-periodic $\Delta_P(\boldsymbol{r})$ modulations signifying a PDW coexisting with superconductivity. Then, by visualizing the temperature dependence of this electronic structure from the superconducting into the pseudogap phase, we find evolution of the scattering interference signature $\Lambda(\boldsymbol{q})$ that is predicted specifically for the temperature dependence of an $8a_0$-periodic PDW. These observations are consistent with theory for the transition from a PDW state coexisting with $d$-wave superconductivity to a pure PDW state in the $Bi_2Sr_2CaDyCu_2O_8$ pseudogap phase.

**Introduction**

Carrier-doped $CuO_2$ sustains both high temperature superconductivity and the pseudogap quantum fluid, often simultaneously. Although the former is reasonably well understood, a decades-long effort by physicists to identify the latter[1,2] has yet to bear fruit. The essential phenomenology of the pseudogap, while complex, is internally consistent. When $p$ holes per unit-cell are introduced to $CuO_2$, the antiferromagnetic insulator (AF) state disappears and the pseudogap (PG) emerges in the region $p<p*$ and $T<T*(p)$ (Fig. 1a). For $T \lesssim T^*(p)$, an energy gap $\Delta^*(p)$ depletes the spectrum of electronic states, and thus the magnetic susceptibility[3] $\chi(T)$, the electronic specific heat[4] $C(T)$, the c-axis conductivity[5,6] $\rho(\omega, T)$, and the average density of electronic states[7] $N(E)$. In $\boldsymbol{k}$-space, there are four $\boldsymbol{k}(E=0)$ Fermi arcs[8] neighboring $\boldsymbol{k} \approx (\pm \pi/2a, \pm \pi/2a)$, beyond which the 'pseudogap' $\Delta^*(\boldsymbol{k})$ opens[3,9,10] near $\boldsymbol{k} \approx (\pm \pi/a, 0); (0, \pm \pi/a)$. At extreme magnetic fields, tiny electron-like pockets with $\boldsymbol{k}$-space area $A_k \approx 7\%$ of the $CuO_2$ Brillouin zone, are detected[11] in the pseudogap state. Probes of electrical and thermal transport in the pseudogap phase evidence electron-pairs without phase rigidity[12-14]. Translational symmetry breaking is widely reported[15-17] to occur within the pseudogap phase; it is associated with charge density modulations of wavevectors $\boldsymbol{Q} \approx 2\pi/a (\pm 1/4, 0); (0, \pm 1/4)$. A $90°$-rotational ($C_4$) symmetry breaking at $\boldsymbol{Q}=\boldsymbol{0}$ and sometimes time-reversal symmetry breaking are also reported depending on materials and technique[18-22]. All these phenomena disappear[10,23,24] near a critical hole density $p = p^*$ which depends on material. The long-term challenge has been to identify a specific state of electronic matter that should exhibit all these properties simultaneously. A viable candidate has emerged recently[25-38], the pair density wave state[39].

A spatially homogeneous $d$-wave superconductor has an electron-pair potential or order parameter $\Delta_d(\boldsymbol{r}) = \Delta_0 e^{i\phi}$ with macroscopic quantum phase $\phi$ and critical temperature $T_c$. By contrast, a PDW state has an order parameter $\Delta_P(\boldsymbol{r})$ that modulates spatially at wavevectors $\boldsymbol{Q}_P$

$$\Delta_P(\boldsymbol{r}) = \left[\Delta(r)e^{i\boldsymbol{Q}_P \cdot \boldsymbol{r}} + \Delta^*(r)e^{-i\boldsymbol{Q}_P \cdot \boldsymbol{r}}\right]e^{i\theta} \tag{1}$$

with a macroscopic quantum phase $\theta$. In theory, such a state exhibits a particle-hole symmetric energy gap $\Delta_P(\boldsymbol{k})$ near the BZ edges, with the $\Delta_P(\boldsymbol{k}) = 0$ points connected by extended $\boldsymbol{k}(E = 0)$ Fermi arcs[8-10]. Of necessity, such a partial gap suppresses $N(E)$, $C(T)$, $\chi(T)$, and $\rho(\omega, T)$. Moreover, a pure PDW is defined by a pair potential modulation as in equation (1) and exhibits a primary electron-pair density modulation $\rho_P(\boldsymbol{r}) = \rho_P^0\left[e^{i2\boldsymbol{Q}_P \cdot \boldsymbol{r}} + e^{-i2\boldsymbol{Q}_P \cdot \boldsymbol{r}}\right]$ along with a collateral charge density modulation $\rho_C(\boldsymbol{r}) = \rho_C^0\left[e^{i\boldsymbol{Q}_C \cdot \boldsymbol{r}} + e^{-i\boldsymbol{Q}_C \cdot \boldsymbol{r}}\right]$ with wavevector $\boldsymbol{Q}_C = 2\boldsymbol{Q}_P$ (Ref. 39). If the PDW is unidirectional, it necessarily breaks the rotation symmetry of the material at $\boldsymbol{Q}$=0, and if biaxial it can break time reversal symmetry[40]. PDW order very naturally produces Fermi arcs[26,30,41,42]. Finally, quasiparticles of the PDW should exhibit scattering interference signatures[35] which are uniquely characteristic of that state.

While charge density modulations $\rho_C(\boldsymbol{r})$ are widely reported in the pseudogap phase[15-17] it is unknown if electron-pair density $\rho_P(\boldsymbol{r})$ or electron-pair potential $\Delta_P(\boldsymbol{r})$ modulations exist therein. Whether the QPI signature $\Lambda_P(\boldsymbol{q})$ of a PDW occurs in the pseudogap phase is also unknown. Indeed, exploration of the pseudogap phase in search of a PDW poses severe experimental challenges. The modulating electron-pair density $\rho_P(\boldsymbol{r})$ which is iconic of the PDW state has been visualized directly by scanned Josephson tunneling microscopy[36,43] but such experiments must be carried out at sub-kelvin temperatures where both sample and STM tip are superconducting. Another approach used in the superconductive phase has been to visualize signatures of the PDW electron-pair potential modulations[35,37,38] $\Delta_P(\boldsymbol{r})$. But none of these experiments provide evidence on whether the pseudogap state in zero magnetic field is a PDW, because they were all carried out deep in the superconducting phase at temperatures $T \lesssim 0.1 T_c$. At low temperatures but in high magnetic fields, both scanning tunneling microscopy and quantum oscillation studies report evidence for a PDW state[37,44], implying that the relict of suppressed superconductivity is a PDW. Therefore, our objective is to visualize the evolution with temperature of electronic structure, especially $\Delta_P(\boldsymbol{r})$ and $\Lambda_P(\boldsymbol{q})$, from the superconducting into the zero-field pseudogap phase of strongly underdoped $Bi_2Sr_2CaDyCu_2O_8$.

## Results

## Modeling the temperature dependence of the PDW state

For theoretical guidance, we use a quantitative, atomic-scale model for PDW state based upon $CuO_2$ electronic structure and the *t-J* Hamiltonian,

$$H = -\sum_{(i,j),\sigma} P_G t_{ij} \left( c_{i\sigma}^\dagger c_{j\sigma} + h.c. \right) P_G + J \sum_{<i,j>} \mathbf{S}_i \cdot \mathbf{S}_j \qquad (2)$$

Here, the electron hopping rates between nearest neighbor (NN) and next-nearest neighbor (NNN) Cu $d_{x^2-y^2}$ orbitals are $t$ and $t'$, respectively, the onsite repulsive energy $U \to \infty$, thus the antiferromagnetic exchange interactions $J=4t^2/U$, and the operator $P_G$ eliminates all doubly-occupied orbitals. A renormalized mean-field theory (RMFT) approximation then replaces $P_G$ with site-specific and bond-specific renormalization factors $g_{i,j}^t$ and $g_{i,j}^s$ based on the average number of charge and spin configurations permissible[34,35]. The resulting Hamiltonian is decoupled into a diagonalizable mean-field approximation using on-site hole density $\delta_i$, bond field $\chi_{ij\sigma}$, and electron-pair potential $\Delta_{ij\sigma}$. This mean field $t$-$J$ Hamiltonian has a uniform $d$-wave superconducting (DSC) state as its ground state, but PDW and DSC states are extremely close in energy, as has also been shown elsewhere[45-47]. Our approach is to find metastable configurations of PDW states and study their signatures in STM. To this end, the RMFT equations are initialized with the electron pair potential fields modulating at wavevector $\boldsymbol{Q}_P = (\pm 1/8,0)2\pi/a_0$, as suggested by recent observations of electron-pair density modulating at $2\boldsymbol{Q}_P$[36] and energy-gap modulations at $\boldsymbol{Q}_P$ at zero-magnetic field[38] as well as in magnetic fields[37]. Moreover because there is little evidence of any long-range magnetic order coexisting with charge modulations in $Bi_2Sr_2CaCu_2O_8$ at any temperatures, we constrain the RMFT solutions to non-magnetic modulating states only, thus, excluding $(\pi,\pi)$ spin density wave order and stripe order. In the self-consistent solution wavefunction $\Psi_0(\boldsymbol{r})$ of this broken-symmetry state then predicts the net charge on each Cu site $\delta_i = 1 - <\Psi_0|\sum_\sigma n_{i\sigma}|\Psi_0>$, the bond-field between adjacent sites $i,j$ $\chi_{ij\sigma} = <\Psi_0|c_{i\sigma}^\dagger c_{j\sigma}|\Psi_0>$, and the electron-pair field on the bond between adjacent sites $i,j$ $\Delta_{ij\sigma} = \sigma<\Psi_0|c_{i\sigma}c_{j\bar\sigma}|\Psi_0>$. Finally, because experimental visualizations are carried out at the crystal termination BiO layer of $Bi_2Sr_2CaCu_2O_8$, Cu $d_{x^2-y^2}$ Wannier functions $W_i(\boldsymbol{r})$ and lattice Green's function $G_{ij\sigma}(E)$ are used to generate the $\boldsymbol{r}$-space Green's functions $G_\sigma(\boldsymbol{r}, E) = \sum_{ij} G_{ij\sigma}(E)W_i(\boldsymbol{r})W_j^*(\boldsymbol{r})$ everywhere at a height 0.4 nm above BiO terminal plane. Thus, the atomically resolved density of electronic states $N(\boldsymbol{r}, E) = \sum_\sigma -\frac{1}{\pi} \text{Im} \, G_\sigma(\boldsymbol{r}, E)$ at the BiO termination surface of $Bi_2Sr_2CaCu_2O_8$ is predicted for the case where the adjacent $CuO_2$ crystal layer sustains a $\lambda = 8a_0$ PDW (Supplementary Note 1).

From this theory, Figure 2a shows the average $N(\boldsymbol{r}, E)$ at height $\sim$4 Å above the BiO termination in $Bi_2Sr_2CaCu_2O_8$ for the PDW state coexisting with d-wave superconductivity (PDW+DSC state) at low-temperatures and pure PDW state at a higher temperature. The PDW+DSC state shows a V-shaped *N(E)* due to presence of nodes in DSC state. With increasing temperature, the uniform component of the pair potential decreases (Supplementary Note 1 and Fig. 2e) and gap scales corresponding to DSC ($\Delta_0$) and PDW ($\Delta_1$) components can be identified as a shoulder feature and a coherence peak, respectively (light-blue curve corresponding to *T*=0.04*t*). Nodal points disappear in transition from PDW+DSC state to PDW state at higher temperatures leading to a large zero-energy *N(E)* in the latter (red curve corresponding to *T*=0.09*t*) (Supplementary Note 1). Thus, a finite zero energy density-of-states is a natural property of a PDW state. These features agree with the experimental findings (Fig. 1d). However, the spectral gap defined by the position of the E>0 coherence peak reduces in the high-temperature PDW state due to the reduced $\Delta_{ij\sigma}$. We believe this discrepancy is a result of an inadequate treatment of self-energy effects in the current renormalized mean field theory, including the assumption of temperature independent Gutzwiller factors (Supplementary Note 1). Fig. 2b shows the most prominent Fourier components of the mean-fields in PDW+DSC and PDW states namely $\boldsymbol{Q}_P = (\pm 2\pi/8\,a_0, 0)$ and $\boldsymbol{Q}_C = 2\boldsymbol{Q}_P$. All mean-fields, including the hole density and the $d$-wave gap order parameter[34] shown in Fig. 2c and 2d, respectively, exhibit periodicity of $8a_0$ in the PDW+DSC state at low temperatures. However, due to the absence of the uniform component ($\boldsymbol{q} = \boldsymbol{0}$) in pure PDW state, density-like quantities are $4a_0$ periodic as predicted by Ginzburg-Landau theories[39] (Fig 2c). We note that the bond field $\chi_{ij}$ is a density-like quantity too and exhibits a behavior very similar to the hole density (Supplementary Note 1). The temperature dependence of the uniform and PDW ($\boldsymbol{q} = \boldsymbol{Q}_P$) components are shown in Fig. 2e. With increasing temperature, the uniform component of the gap, which corresponds to DSC in the PDW+DSC state, decreases rapidly and becomes negligibly small compared to the PDW component in the temperature range $0.05t < T < 0.085t$, but does not vanish. We have verified that a converged nonzero solution for $\Delta(\boldsymbol{q} = \boldsymbol{0})$, 'fragile PDW+DSC' state', exists in this region (white background in Figs. 2e-f). For $T > 0.085t$, the PDW+DSC solution of the RMFT equations becomes unstable and the pure PDW state is the only stable solution for a modulated state (pink background in Figs. 2e-f) (Supplementary Note 1). The temperature dependence of $\boldsymbol{q} = \boldsymbol{Q}_P$ and $\boldsymbol{q} = \boldsymbol{Q}_C$ components of the hole density is shown in Fig. 2f. We find that the $\boldsymbol{q} = \boldsymbol{Q}_C$ component of the charge density is dominant at all temperatures and the $\boldsymbol{q} = \boldsymbol{Q}_P$ component exhibits essentially the same temperature dependence as the uniform component of the gap order parameter. This is in agreement with Ginzburg–Landau theory[39] and experimental observation[37,38,43] that a PDW generated charge density wave (CDW) state will have $\boldsymbol{q} = \boldsymbol{Q}_P$ and $\boldsymbol{q} = \boldsymbol{Q}_C = 2\boldsymbol{Q}_P$ components that are related to the uniform ($\Delta(\boldsymbol{0})$) and PDW ($\Delta(\boldsymbol{Q}_P)$) components of the gap order parameter as $\delta(\boldsymbol{Q}_P) \propto$ $(\Delta(\mathbf{0})\Delta(-\boldsymbol{Q}_P)^* + \Delta(\boldsymbol{Q}_P)\Delta(\mathbf{0})^*)$ and $\delta(\boldsymbol{Q}_C) \propto \Delta(\boldsymbol{Q}_P)\Delta(-\boldsymbol{Q}_P)^*$. The self-consistent PDW solutions are found to exist in a hole doping range 0.06<$p$<0.14 for all temperatures considered.

We explore these predictions using strongly underdoped Bi$_2$Sr$_2$CaDyCu$_2$O$_8$ samples with resistive transition temperature $T_c = 37 \pm 3$ K and $p \cong 0.08$ as shown schematically by the white arrow in Fig. 1a. These samples are cleaved in cryogenic vacuum at $T \approx 4.2$ K and inserted to the instrument. Measurements are carried out at a sequence of temperatures from $0.1T_c \leq T \leq 1.5T_c$ spanning the range from the superconducting to well into the pseudogap phase. The topographic images $T(\boldsymbol{r})$ of the FOV studied versus temperature are taken using the experimental methods described in "Methods" section and presented in Supplementary Figure 4. Both the tip-sample differential tunneling current $I(\boldsymbol{r}, V)$ and conductance $dI/dV(\boldsymbol{r}, E = eV) \equiv g(\boldsymbol{r}, V)$ are measured at bias voltage $V=E/e$ and with sub-angstrom spatial resolution. Because the density-of-electronic-states $N(\boldsymbol{r}, E)$ is related to the differential conductance as $g(\boldsymbol{r}, E) \propto N(\boldsymbol{r}, E)[I_s / \int_0^{eV_s} N(\boldsymbol{r}, E')dE']$, where $I_s$ and $V_s$ are arbitrary set-point parameters and the denominator $\int_0^{eV_s} N(\boldsymbol{r}, E')dE'$ is unknown, valid imaging of $N(\boldsymbol{r}, E)$ is intractable. However, one can suppress these serious systematic "set-point" errors by using $R(\boldsymbol{r}, E) = I(\boldsymbol{r}, E)/I(\boldsymbol{r}, -E)$ or $Z(\boldsymbol{r}, E) = g(\boldsymbol{r}, E)/g(\boldsymbol{r}, -E)$ so that distances, modulation wavelengths and spatial-phases can be measured accurately. Furthermore, Bogoliubov quasiparticle scattering interference (BQPI) occurs when an impurity atom scatters quasiparticles, which interfere to produce characteristic modulations of $\delta N(\boldsymbol{r}, E)$ surrounding each scattering site. The Fourier transform of $\delta N(\boldsymbol{r}, E)$, $\delta N(\boldsymbol{q}, E)$, then exhibits intensity maxima at a set of wavevectors $\boldsymbol{q}_i$ connecting regions of high joint-density-of-states. Local maxima in $Z(\boldsymbol{q}, E)$ therefore reveal the sets of energy dispersive wavevectors $\boldsymbol{q}_i(E)$ generated by the scattering interference. An efficient synopsis of these complex phenomena can then be achieved[10] by using $\Lambda(\boldsymbol{q}, \Delta) = \sum_{E \cong 0}^{\Delta} Z(\boldsymbol{q}, E)$, which provides a characteristic "fingerprint" of whatever ordered state, e.g. CDW or PDW, controls the $\boldsymbol{q}_i(E)$.

**Evolution of energy gap modulations from superconductive to pseudogap phase**

At $T$ = 4.2 K we first measure $g(\boldsymbol{r}, V)$ in a 20 nm-square FOV (see "Methods" section and Supplementary Figure 4). The average differential conductance $g(V)$ is shown as a blue curve in Fig. 1d, where the energy of the coherence peak is determined from a local maximum in $g(V)$ for $V > 0$ (identified by a black vertical arrow). Measuring this energy versus location yields the so-called gapmap $\Delta_1(\boldsymbol{r})$ as shown in Fig. 3a. Fourier analysis of $\Delta_1(\boldsymbol{r})$ yields $\Delta_1(\boldsymbol{q})$, which exhibits significant disorder as $\boldsymbol{q} \to \boldsymbol{0}$ (Supplementary Figure 5b). But, by fitting the central peak to a cylindrical gaussian, and then subtracting it from $\Delta_1(\boldsymbol{q})$, we find four maxima at $\boldsymbol{q} \approx [(\pm 0.125 \pm 0.040, 0); (0, \pm 0.125 \pm 0.015)] \, 2\pi/a_0$ (inset in Fig 3a). These are the energy-gap modulations with period approximately $8a_0$, that have been previously reported[35, 37,38] for samples with $p \approx 0.17$, and are the signature of a PDW state coexisting with $d$-symmetry superconductivity at low temperature. Fourier filtration of Fig. 3a retaining only modulations at $\boldsymbol{q} \approx [(\pm 1/8, 0); (0, \pm 1/8)] \, 2\pi/a_0$ yields an accurate image of the PDW gap modulations as seen in Fig. 3b. But when the same procedures are carried out at $T = 1.5T_C = 55$ K, the coherence peaks from which the gap is defined have so diminished that an equivalent gapmap is difficult to achieve. For example, Figs 3c and 3d show the measured $g(\boldsymbol{r}, 60 \, mV)$ in an identical 10 nm-square FOV at $T = 0.14T_C = 5$ K and $T = 1.5T_C = 55$ K. Cross correlation analysis of $g(\boldsymbol{r}, V)$ at $T = 0.14T_C$ and of $g(\boldsymbol{r}, V)$ at $T = 1.5T_C$ in this FOV versus bias voltage V, yield a normalized cross correlation coefficients around 0.9 for practically all energies (Supplementary Note 3), thus indicating that virtually no changes have occurred in spatial arrangements of electronic structure upon entering the PG phase. The major exception is in the energy range $+100$ meV $< E < +160$ meV, wherein the feature denoted coherence peak (arrow Fig. 1d) diminishes strongly in amplitude. This however makes comparison of the $\Delta_1(\boldsymbol{r})$ in same FOV at $T = 5$ K and $T = 55$ K challenging. Figures 3e and 3f show the measured $\Delta_1(\boldsymbol{r})$ and, where it is possible to determine the energy, no changes have occurred in spatial arrangements of energy gaps either. The cross-correlation coefficient between the $\Delta_1(\boldsymbol{r}, 0.14T_C)$ and $\Delta_1(\boldsymbol{r}, 1.5T_C)$ is 0.685 indicating that the PDW state found at $T \ll T_C$ remains robustly present at $1.5T_C$ deep into the pseudogap phase.

**Temperature Evolution in the QPI Signature of a PDW State**

Next, we measure $g(\boldsymbol{r}, V, T)$ for $-34$ mV $< V < 34$ mV at a sequence of temperatures $0.1T_c \leq T \leq 1.5T_c$. Then $Z(\boldsymbol{r}, V) = g(\boldsymbol{r}, +V)/g(\boldsymbol{r}, -V)$ is evaluated for each temperature, and the power-spectral-density Fourier transforms $Z(\boldsymbol{q}, V)$ are derived. Hence, $\Lambda(\boldsymbol{q}, \Delta_0) = \sum_{E \cong 0}^{\Delta_0} Z(\boldsymbol{q}, E)$ is calculated at each temperature where $\Delta_0 = 20$ meV is the observed energy above which dispersive scattering interference is no longer detectable[10,35]. The measured temperature dependence of $\Lambda(\boldsymbol{q}, \Delta_0)$ for $0.1T_c \leq T \leq 1.5T_c$ is shown in the left column in Fig. 4. The initial $\Lambda(\boldsymbol{q}, \Delta_0)$ features at $T=4.2$K are exactly as expected from theory and as observed by experiment at $p = 0.17$, for a PDW coexisting with a $d$-wave superconductor[14]. As temperature increases the characteristics remain strikingly unchanged except that the intensity become significantly weaker. That the passage through $T_c$ exhibits almost no signature in $\Lambda(\boldsymbol{q}, \Delta_0)$, is unexpected if the scattering interference in $\Lambda(\boldsymbol{q}, \Delta_0)$ is only due to the $d$-wave superconductivity. If however, a PDW state exists both below and above $T_c$ this is what might logically be expected. Moreover, quantitative theoretical predictions for $\Lambda(\boldsymbol{q}, \Delta_0)$ for a $\lambda = 8a_0$ PDW using the RMFT model, predict $Z(\boldsymbol{q}, E)$ surrounding a point-like scatterer and hence $\Lambda_P(\boldsymbol{q}, \Delta_0) = \sum_{E \cong 0}^{\Delta_0} Z(\boldsymbol{q}, E)$ (Supplementary Note 5). Comparing our $\Lambda(\boldsymbol{q}, \Delta_0)$ data to the RMFT-derived predictions $\Lambda_P(\boldsymbol{q}, \Delta_0)$ for a $\lambda = 8a_0$ PDW in the right column in Fig. 4 shows how the key features in the experiments are reproduced in the theory over the whole range of temperatures. Features at $\boldsymbol{q} \approx (\pm 1/4, \pm 1/4)\ 2\pi/a_0$ extending in nodal directions disappear with the transition from superconducting state to pseudogap; this is reproduced in our theory as a consequence of vanishing DSC component (indicated by red arrow in Fig. 4 and Supplementary Figure 12). Further, the measured length of the $\Lambda(\boldsymbol{q}, \Delta_0)$ arc features about $(\pm 1, \pm 1)\ 2\pi/a_0$ increase continuously from superconducting to pseudogap phase (Supplementary Note 6). This temperature evolution of the arc length in $\Lambda_P(\boldsymbol{q}, 20\text{ meV})$ from PDW+DSC state to pure PDW state in the RMFT model, has indistinguishable characteristics (Supplementary Figure 14). Moreover, superimposing the experimental and theoretical maps shows nearly identical positioning of dominant QPI features in q-space (Supplementary Figure 12). The implication is that the PDW state which definitely exists at lowest temperatures[35-38], continues to exist into pseudogap phase. But in that case, since that pseudogap is often (but not always) reported to support no supercurrents, it would have to be in a strongly phase fluctuating PDW phase [32,33,50-54].

## Comparison of QPI Signature of a CDW and PDW State

Finally, we consider the widely promulgated hypothesis[15,16,17,] that the pseudogap phase is a primary CDW state, whose charge density modulation breaks the translational symmetry of the cuprate pseudogap phase. First we note the very sharp distinction between these states: the mean-field order parameter of a PDW at wavevector $\boldsymbol{Q}_P$ is $\langle c^\dagger_{k\uparrow} c^\dagger_{-k+Q_P\downarrow} \rangle$, whereas for a CDW at wavevector $\boldsymbol{Q}_C$ it is $\sum_\sigma \langle c^\dagger_{k,\sigma} c_{k+Q_C,\sigma} \rangle$. Second, while a periodically modulating energy gap is a key PDW signature (Fig. 3a), $\boldsymbol{r}$-space energy gap modulation should be weak in a CDW state, where it is charge density which modulates. Third, the quasiparticles and their scattering interference are highly distinct for the two states. A primary CDW order by itself does not exist as a stable self-consistent solution of the RMFT $t$-$J$ model at any temperatures or dopings that we have considered. However, we can study STM signatures of the CDW order non-self-consistently. Figure 5a shows the predicted $\Lambda_P(\boldsymbol{q}, \Delta_0) = \sum_{E\cong 0}^{\Delta_0} Z(\boldsymbol{q}, E)$ for a $\lambda = 8a_0$ PDW in the CuO$_2$ pseudogap phase, while Fig. 5b shows the equivalent predictions for a $\lambda = 4a_0$ CDW (Supplementary Note 5). In Fig. 5c, d we show the measured $\Lambda(\boldsymbol{q}, \Delta_0)$ at $T = 1.25T_c$ and $T = 1.5T_c$ ($\Lambda(\boldsymbol{q}, \Delta_0)$ analysis details are discussed in Supplementary Note 6). Clearly, the measured $\Lambda(\boldsymbol{q}, \Delta_0)$ is in superior agreement with the $\Lambda_P(\boldsymbol{q}, \Delta_0)$ signature of a $\lambda = 8a_0$ PDW rather than with that of a $\lambda = 4a_0$ CDW. The energy evolution of the wavevectors is visualized in the measured $Z(\boldsymbol{q}, V, 55 \text{ K})$ from 4 mV to 20 mV (movie S2 and Supplementary Note 7). The wavevectors evolve dispersively with energy only by a small amount. Finally, The $\Lambda(\boldsymbol{q}, \Delta_0)$ in the pseudogap phase forms an open contour near the lines $(\pm 1, 0)2\pi/a_0$ and $(0, \pm 1)2\pi/a_0$; this is consistent with the open contours in the $\Lambda_P(\boldsymbol{q}, \Delta_0)$ signature of a $\lambda = 8a_0$ PDW but distinct from the closed contours of a $\lambda = 4a_0$ CDW. Therefore, predictions of a pure PDW theory corresponds well and in detail to the complex patterns of the quasiparticle scattering that are actually observed in the pseudogap phase of Bi$_2$Sr$_2$CaDyCu$_2$O$_8$.

To summarize: strongly underdoped $Bi_2Sr_2CaDyCu_2O_8$ at $p\sim0.08$ and $T = 5$ K exhibits the $8a_0$-periodic $\Delta_1(\boldsymbol{r})$ modulations characteristics of a PDW coexisting with superconductivity[35,37,38] (Fig. 2d, Fig 3b). Increasing temperature from the superconducting into the pseudogap phase, seems to retain these real-space phenomena apparently thermally broadened but otherwise unchanged (Fig. 3c-f). More obviously, the measured scattering interference signature[10] $\Lambda(\boldsymbol{q})$ evolves from correspondence with $\Lambda_P(\boldsymbol{q})$ predicted for an $8a_0$-periodic PDW coexisting with superconductivity[35] into that predicted for a pure $8a_0$-periodic PDW above the superconductive $T_c$ in the pseudogap phase (Fig. 4). Furthermore, this signature is highly distinct from $\Lambda(\boldsymbol{q})$ predicted for a $4a_0$-periodic CDW (Fig. 5). The clear inference from all these observations is that the $Bi_2Sr_2CaDyCu_2O_8$ pseudogap phase contains a PDW state, whose quantum phase is fluctuation dominated.

## Methods

Single crystals of $Bi_2Sr_2CaDyCu_2O_8$ with hole doping level of $p \approx 8\%$ and $T_c = 37\pm3$ K were synthesized using the floating zone method with doping controlled by oxygen depletion. The samples were cleaved in cryogenic ultrahigh vacuum at $T = 4.2$ K to reveal an atomically flat BiO surface, and then inserted into STM. All measurements are carried out using tungsten tips in a variable temperature (the range is $T = 4.2$ K – 55 K) spectroscopic imaging STM system with thermal fluctuations less than 1 mK. The PG gap map $\Delta_1(\boldsymbol{r})$ were measured with the resolution of 128 pixels × 128 pixels. The experimental parameters include setpoint voltage 800 mV, setpoint current 800 pA, bias voltage $V_B$ = -800 mV – 800 mV and 161 discrete energy layers. The QPI images were measured with the resolution of 256 pixels × 256 pixels. The experimental parameters of the QPI measurements include spectroscopic setpoint voltage 200 mV, setpoint current 200 pA, bias voltage $V_B$ = -34 mV – 34 mV and 35 discrete energy layers. The topography $T(\boldsymbol{r})$ of the six temperatures studied in this paper are shown in Supplementary Figure 4. The presented QPI patterns were symmetrized to reduce the noise. In the QPI pattern, a circle with a locus located at $\boldsymbol{q} = \boldsymbol{0}$ and a radius of 25 pixels is fitted to 2D Gaussian function and then removed.

**Acknowledgements:** The authors acknowledge and thank K. Fujita for very helpful discussions and advice. Y.X.C. and J.C.S.D acknowledge the Moore Foundation's EPiQS Initiative through Grant GBMF9457. J.C.S.D. acknowledge support from Science Foundation of Ireland under Award SFI 17/RP/5445. W.C. and J.C.S.D. acknowledge support Royal Society under Award R64897. S.W., W.R. and J.C.S.D. acknowledge support from the European Research Council (ERC) under Award DLV-788932. S.W. acknowledges support from John Fell Fund at Oxford University. P.J.H. acknowledges support from research grant NSF-DMR-1849751.

**Competing Interests:** The authors declare no competing interests.

**Data availability**

All data are available in the main text, in the Supplementary Information and on Zenodo[55]. Additional information is available from the corresponding author upon reasonable request.

**Code availability**

The data analysis codes used in this study are available from the corresponding author upon reasonable request.

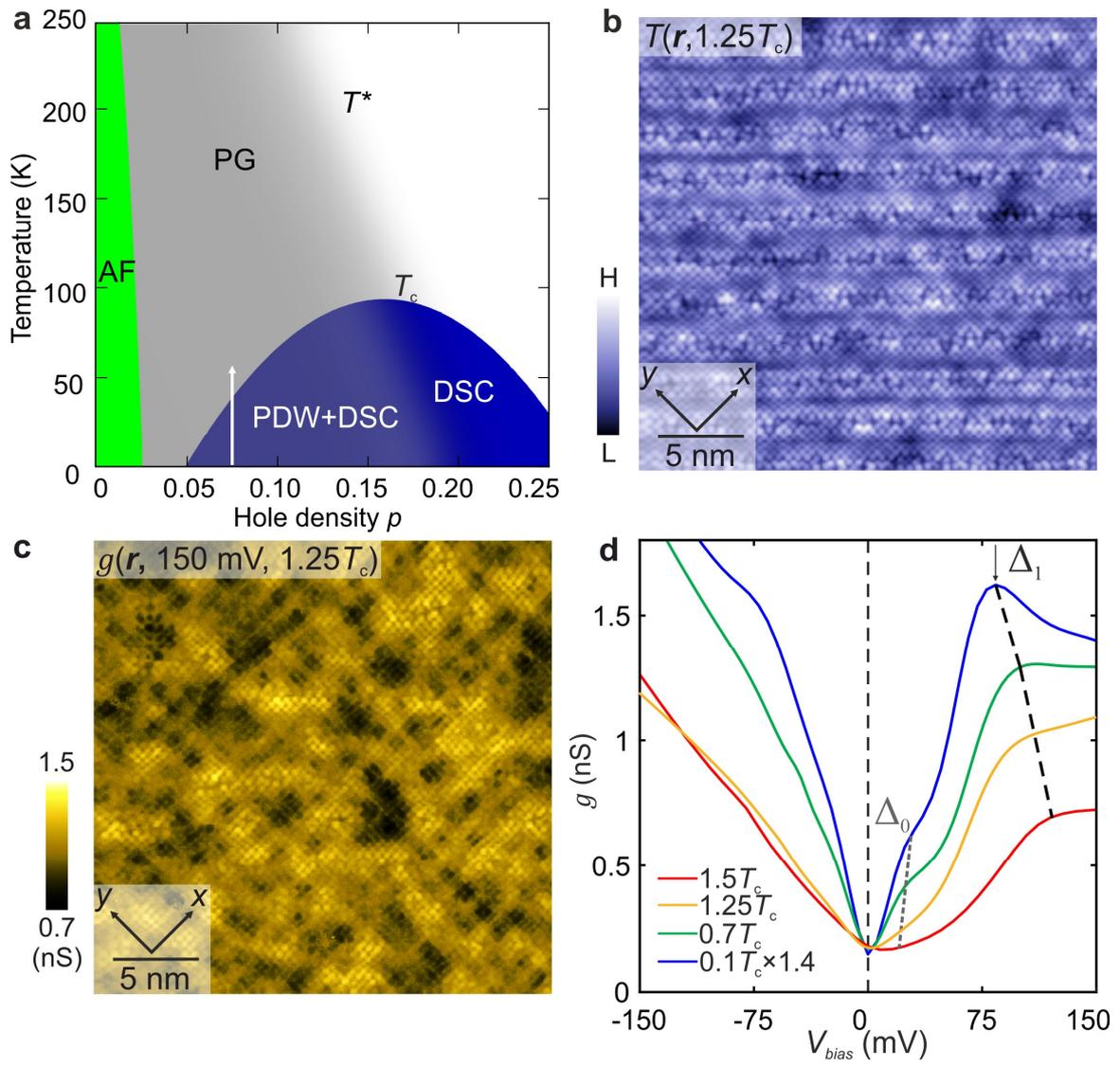

**a**

Temperature (K), Hole density $p$, AF, PG, $T^*$, $T_c$, DSC, PDW+DSC

**b** $T(\boldsymbol{r}, 1.25T_c)$

H, L, 5 nm

**c** $g(\boldsymbol{r}, 150$ mV, $1.25T_c)$

1.5, 0.7 (nS), 5 nm

**d**

$g$ (nS), $V_{bias}$ (mV), $\Delta_1$, $\Delta_0$

1.5$T_c$
1.25$T_c$
0.7$T_c$
0.1$T_c \times 1.4$

**FIG. 1 Temperature dependence of cuprate broken-symmetry states.**

**a.** Schematic phase diagram of hole-doped cuprates. The Mott insulator phase with long range antiferromagnetic order (AF) is replaced by the pseudogap phase (PG) with increasing hole doping $p$ below the onset temperature $T^*$. The PG phase is characterized by the suppression of magnetic susceptibility, electronic specific heat, the c-axis conductivity and the average density of electronic states, and the appearance of a truncated Fermi surface. The d-symmetry Cooper-paired high-temperature superconductivity state (DSC) is indicated schematically in a blue "dome". The range of temperature $T$, in which the PG state is studied in this paper is indicated by the white arrow.

**b.** Topograph $T(\boldsymbol{r})$ at the BiO termination layer at $T = 1.25T_c$ in the PG phase of $Bi_2Sr_2CaDyCu_2O_8$ for $p \approx 0.08$.

**c.** Differential conductance map $g(\boldsymbol{r}, +150\ mV)$ was obtained at the same field of view as B at $T = 1.25T_c = 45$ K. The $g(\boldsymbol{r}, E)$ manifests $\lambda = 4a_0$ charge modulations.

**d.** Evolution of the spatially averaged tunneling conductance spectra of $Bi_2Sr_2CaDyCu_2O_8$ with increasing $T$, here characterized by $T_c$. The gap $\Delta_1(T)$ is the energy of the coherence peak that is identified by a local maximum in $g(V)$ for V>0 (indicated by a black vertical arrow). The energies $\Delta_0(T)$ (gray dashed line) are identified as the extinction energy of Bogoliubov quasiparticles (see movie S1). The two characteristic energies $\Delta_0(T)$ and $\Delta_1(T)$ appear more subtle at higher temperatures due to thermal broadening. Note the tunneling spectra at 4.2 K ($\approx 0.1T_c$) is multiplied by 1.4.

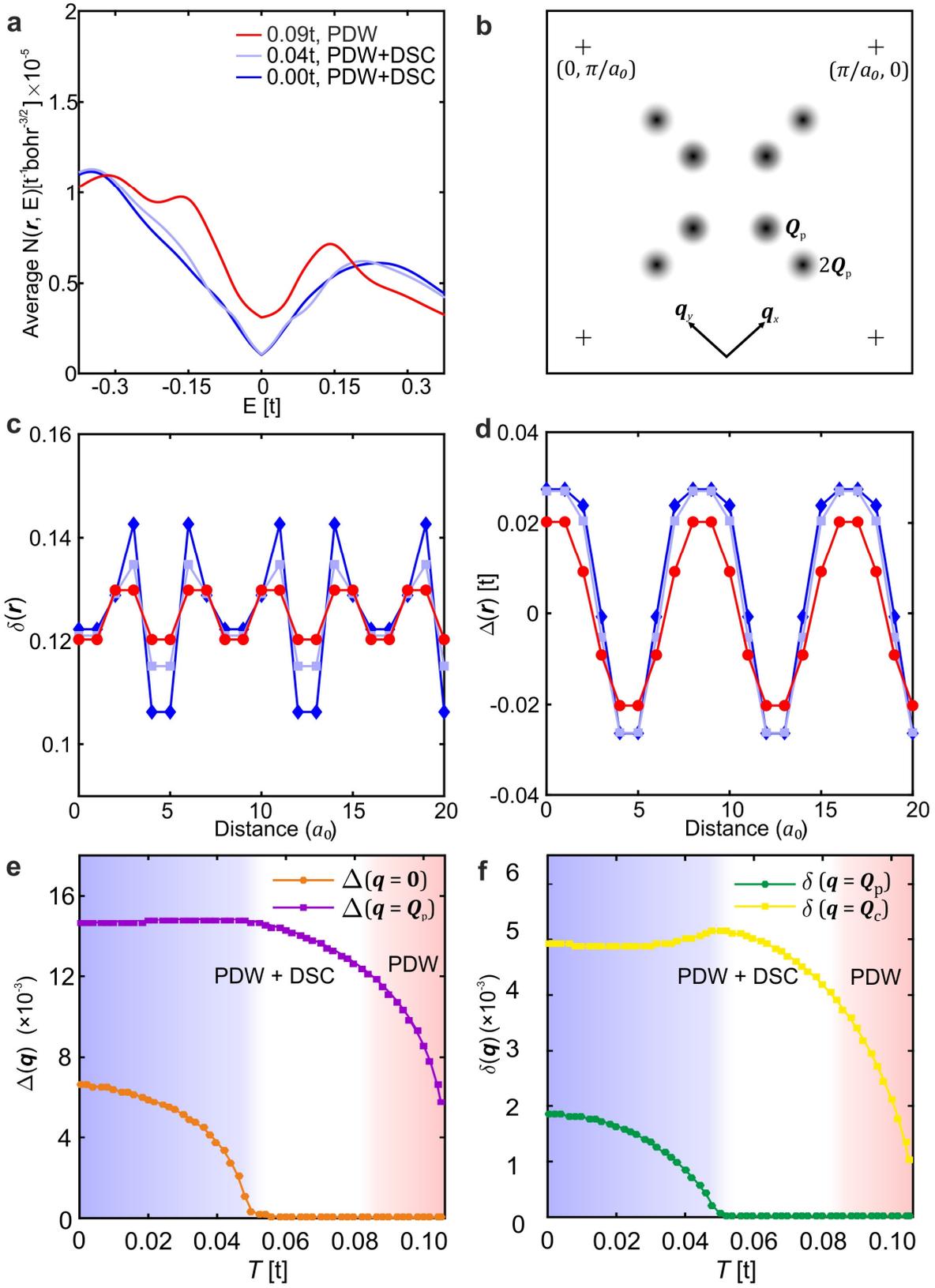

**Fig. 2 Predicted temperature-evolution of the average local density of states ($N(\boldsymbol{r})$), hole density ($\delta$) and d-wave gap order parameter ($\Delta$) for the PDW state.**

**a.** Continuum LDOS $N(\boldsymbol{r}, \text{E})$ spatially averaged over a period of PDW ($8a_0$) at low-temperatures ($T$=0 and $T$=0.04$t$) and pure PDW state at a high temperature ($T$=0.09$t$) obtained using parameter set doping $p$ = 0.125, $t$ = 400 meV, $t'$ = -0.3$t$ and $J$ = 0.3$t$. PDW+DSC state exhibits V-shaped nodal LDOS due to presence of the DSC component. Pure PDW state has Bogoliubov-Fermi pockets (in contrast to nodes in PDW+DSC state), which leads to a large E = 0 LDOS. The LDOS is obtained after the effects of linear inelastic scattering $i0^+ + i\Gamma$ is incorporated, where $\Gamma = \alpha|E|$ and $\alpha$ = 0.25 using the experimental fits in Ref. [48]. A non-zero LDOS at zero-bias in PDW+DSC state is a consequence of the finite artificial broadening $i0^+$.

**b.** q-space schematic showing the most prominent wavevectors $\boldsymbol{Q}_P$ = [(±1/8, 0); (0, ±1/8)]2π/$a_0$ and 2$\boldsymbol{Q}_P$ appearing in the Fourier transform of the mean-fields and other related quantities.

**c.** Spatial variation of hole density ($\delta$) in PDW+DSC state (at $T$=0 and $T$=0.04$t$) and pure PDW state (at $T$=0.09$t$). Hole density modulates with a periodicity of $8a_0$ in PDW+DSC state due to presence of the DSC component and a periodicity of $4a_0$ in pure PDW state due to the absence of the DSC component, as expected from Ginzburg-Landau theories.

**d.** Spatial variation of d-wave gap order parameter in PDW+DSC state (at $T$=0 and T=0.04$t$) and pure PDW state (at $T$=0.09$t$) exhibiting $8a_0$-periodic modulations corresponding to the PDW component of the gap.

**e.** Temperature evolution of the uniform ($\boldsymbol{q} = \boldsymbol{0}$) and PDW ($\boldsymbol{q} = \boldsymbol{Q}_P$) components of the d-wave gap order parameter in PDW+DSC state (0<$T$<0.085t) and pure PDW state (0.085$t$<$T$<0.11$t$). The uniform component of the gap decreases sharply with temperature becoming negligibly small, but finite, compared to the PDW component for 0.05$t$<$T$<0.085$t$. This 'fragile PDW+DSC' state is shown in white background. For $T$>0.085$t$ PDW+DSC state becomes unstable and only pure PDW state (shown in pink background) exists as a stable solution of the RMFT equations.

**f.** Temperature evolution of the $\boldsymbol{q} = \boldsymbol{Q}_P$ and $\boldsymbol{q} = \boldsymbol{Q}_C = 2\boldsymbol{Q}_P$ components of hole density ($\delta$) in PDW+DSC and pure PDW state in the same temperature range as in (**e**). The $\boldsymbol{q} = \boldsymbol{Q}_P$ component mirrors the temperature evolution of the uniform component of the gap (panel (**e**), as expected from Ginzburg-Landau theories. $\boldsymbol{q} = \boldsymbol{Q}_C$ component is the dominant component at all temperature leading to $4a_0$–periodic charge density wave.

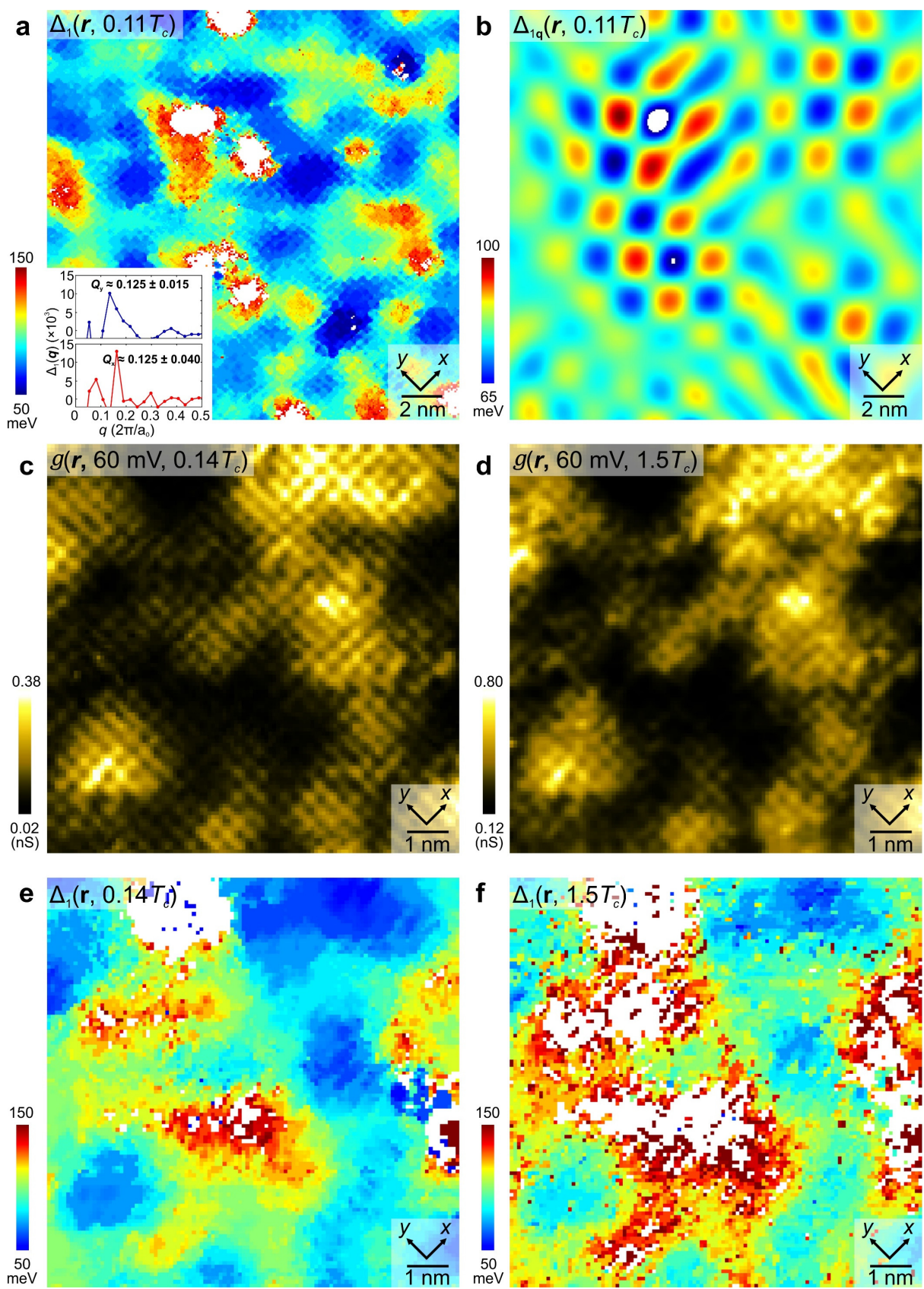

**a** $\Delta_1(\mathbf{r}, 0.11\,T_c)$

$\mathbf{Q}_x \approx 0.125 \pm 0.015$

$\mathbf{Q}_y = 0.125 \pm 0.040$

$\Delta_1(\mathbf{q})\,(\times 10^3)$

$q\,(2\pi/a_0)$

150
50
meV

100
65
meV

**b** $\Delta_{1\mathbf{q}}(\mathbf{r}, 0.11\,T_c)$

**c** $g(\mathbf{r}, 60\,\text{mV}, 0.14\,T_c)$

0.38
0.02
(nS)

**d** $g(\mathbf{r}, 60\,\text{mV}, 1.5\,T_c)$

0.80
0.12
(nS)

**e** $\Delta_1(\mathbf{r}, 0.14\,T_c)$

150
50
meV

**f** $\Delta_1(\mathbf{r}, 1.5\,T_c)$

150
50
meV

**FIG 3. Energy-gap Modulations from Superconductive to Pseudogap Phase.**

**a.** Measured $\Delta_1(\boldsymbol{r})$ within 20 nm × 20 nm FOV at $T$= $0.11T_C$ = 4.2K. The energy-gap is measured from the energy of the coherence peak at E > 0. The inset shows the linecuts from $\boldsymbol{q}$ = (0, 0) to (0.5, 0)2π/$a_0$ and from $\boldsymbol{q}$ = (0, 0) to (0, 0.5)2π/$a_0$ in the measured $\Delta_1(\boldsymbol{q})$ after subtraction of the disorder core. $\boldsymbol{q} \approx [(\pm 1/8, 0); (0, \pm 1/8)]$ 2π/$a_0$ peaks are present in both directions. The white areas represent regions where it is impossible to determine the coherence peak position $\Delta_1$.

**b.** Gap modulations $\Delta_{1q}(\boldsymbol{r})$ from 3(**a**). these are visualized at wavevectors $\boldsymbol{q} \approx [(\pm 1/8, 0); (0, \pm 1/8)]$ 2π/$a_0$ by Fourier filtering $\Delta_1(\boldsymbol{r})$ at the 1/8 peaks as shown in inset of 3a. The Gaussian filter size $\sigma_q$ = 1.45 pixels (or equivalently 0.455 nm$^{-1}$) in $\boldsymbol{q}$-space, which corresponds to 2.2 nm in $\boldsymbol{r}$-space.

**c.** Measured $g(\boldsymbol{r}, 60\ mV)$ at $T$ = $0.14T_C$ = 5 K within 9.9 nm × 9.9 nm FOV. The $g(\boldsymbol{r}, 60\ mV)$ manifests unidirectional charge modulations.

**d.** Measured $g(\boldsymbol{r}, 60\ mV)$ at $T$ = $1.5T_C$ = 55 K in the identical FOV as (**c**). No change has been detected in $g(\boldsymbol{r}, 60\ mV)$ at $T$ = 55 K.

**e.** Measured $\Delta_1(\boldsymbol{r})$ at $T$ = $0.14T_C$ = 5 K shows the spatial variation of the coherence peak at E > 0.

**f.** Measured $\Delta_1(\boldsymbol{r})$ at $T$ = $1.5T_C$ = 55 K in the identical FOV as (**c**), (**d**) and (**e**). The spatial variation of the coherence peak is highly similar to (**e**).

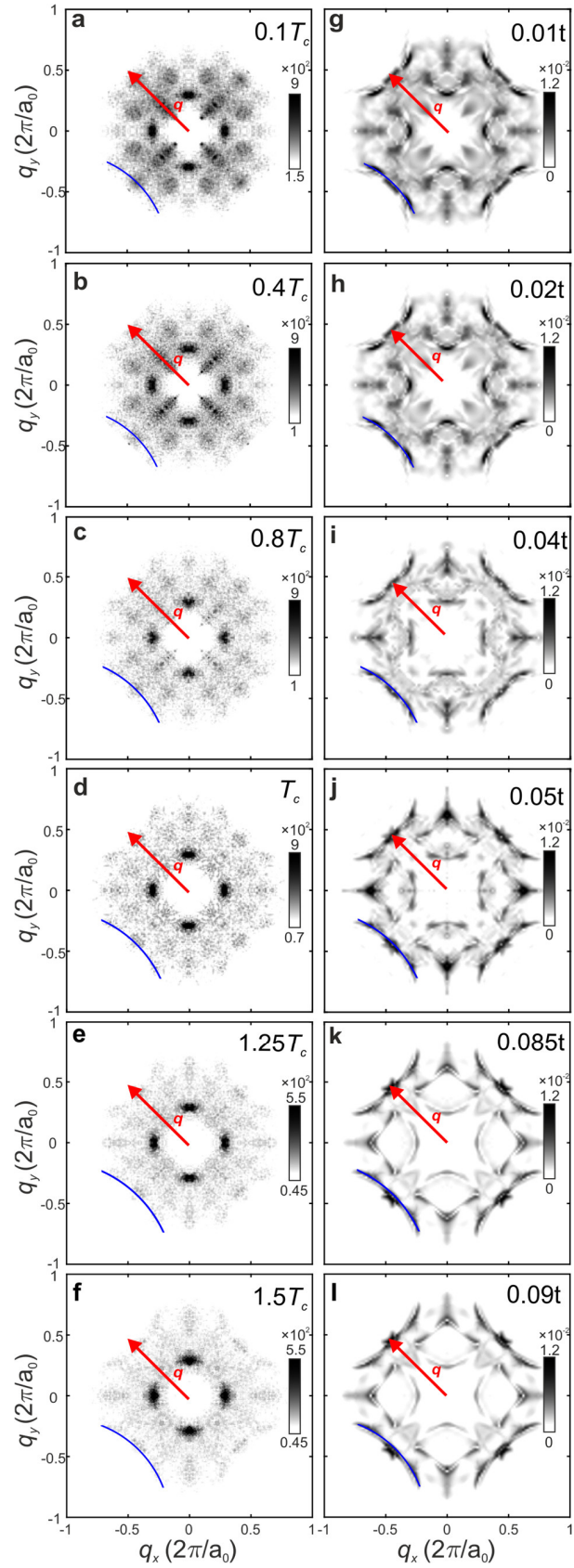

**FIG 4. Temperature Dependence of QPI Signature of a PDW**

**a-f**. Measured QPI signature $\Lambda(\boldsymbol{q}, 20 \text{ meV})$ for $Bi_2Sr_2CaDyCu_2O_8$ (doping level $p \approx 0.08$) at temperatures $T$ = (**a**) $0.1T_c$, (**b**) $0.4T_c$, (**c**) $0.8T_c$, (**d**) $T_c$, (**e**) $1.25T_c$, and (**f**) $1.5T_c$.

**g-j**. Predicted QPI signature $\Lambda_P(\boldsymbol{q}, 20 \text{ meV})$ of $8a_0$ PDW state that coexists with DSC state at temperatures $T$ = (**g**) $0.01t$, (**h**) $0.02t$, (**i**) $0.04t$, and (**j**) $0.05t$. Theoretically, it is assumed that the short-range discommensurate nature of the charge order, as seen in the experiments[49], will lead to reduced intensity of the density wave Bragg peaks compared to the long-range PDW driven charge order obtained in our mean-field analysis. Accordingly, the non-dispersing charge order Bragg peaks at wavevectors $\boldsymbol{q} = \pm n\boldsymbol{Q}_p$, $n$ = 0, 1, 2, ...,7, in PDW+DSC state and $\boldsymbol{q} = \pm n(2\boldsymbol{Q}_p)$, $n$ = 0, 1, 2, 3, in the pure PDW state are suppressed by a factor of 100 in $\Lambda_P(\boldsymbol{q}, 20 \text{ meV})$, which helps in highlighting much weaker wavevectors emerging from impurity scattering. $\Lambda_P(\boldsymbol{q}, 20 \text{ meV})$ is computed for unidirectional PDW in a 56×56 lattice and symmetrized for plotting. Features at $\boldsymbol{q} \approx (\pm 1/4, \pm 1/4) \, 2\pi/a_0$ extending in nodal directions are labeled by a red arrow.

**k-l**. Predicted $\Lambda_P(\boldsymbol{q}, 20 \text{ meV})$ of pure $8a_0$ PDW state at temperatures $T$ = (**k**) $0.085t$ and (**l**) $0.09t$. Measured $\Lambda(\boldsymbol{q}, 20 \text{ meV})$ in (**a-f**) for $T = 0.1T_c \sim 1.5T_c$ are in good agreement with the simulation results in (**g-l**). The length of the arc-like feature (indicated by blue curves) near $(\pm 1, \pm 1) \, 2\pi/a_0$ increases from PDW+DSC to pure PDW state, which is a key feature of charge order driven by PDW. The intensity of $\Lambda(\boldsymbol{q}, 20 \text{ meV})$ and $\Lambda_P(\boldsymbol{q}, 20 \text{ meV})$ decreases as the temperature increases.

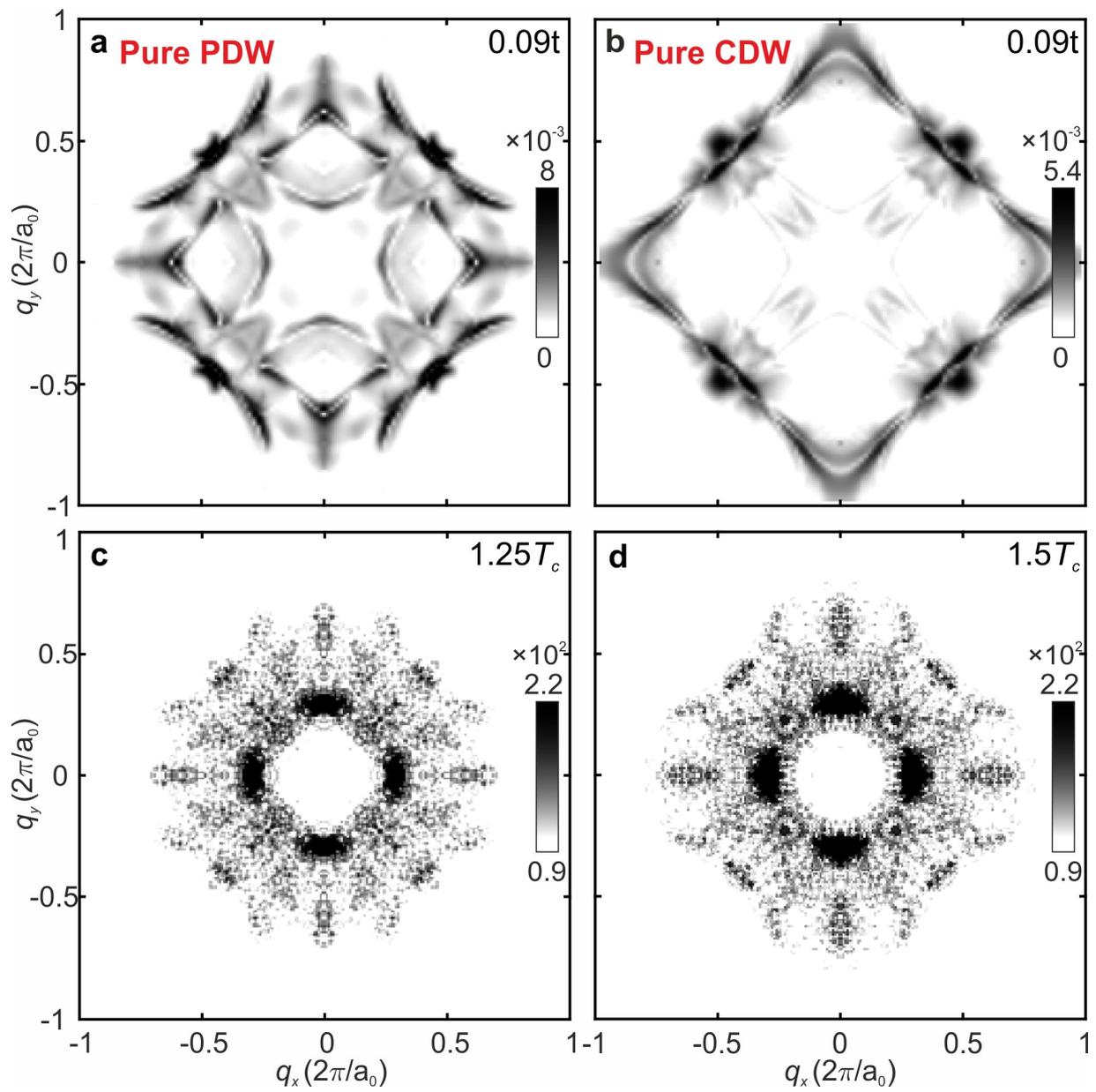

**FIG 5. Discrimination of CDW from PDW QPI Signature in the Pseudogap State**

**a.** Predicted QPI signature $\Lambda_P(\boldsymbol{q}, 20\text{ meV})$ of pure $8a_0$ PDW state at $T = 0.09t$.

**b.** Predicted $\Lambda_C(\boldsymbol{q}, 20\text{ meV})$ of pure $4a_0$ CDW state at $T = 0.09t$. The CDW states show very different features compared to the PDW state.

**c.** Measured $\Lambda(\boldsymbol{q}, 20\text{ meV})$ of $Bi_2Sr_2CaDyCu_2O_8$ ($p \approx 0.08$) for the pseudogap phase at $T = 1.25T_c$.

**d.** Measured $\Lambda(\boldsymbol{q}, 20\text{ meV})$ of $Bi_2Sr_2CaDyCu_2O_8$ ($p \approx 0.08$) for the pseudogap phase at $T = 1.5T_c$. The measurements of the pseudogap phase agree much better with the pure PDW scenario (**a**) than with the pure CDW (**b**).

Supplementary Information for

# Scattering Interference Signature of a
# Pair Density Wave State
# in the Cuprate Pseudogap Phase

Shuqiu Wang, Peayush Choubey, Yi Xue Chong, Weijiong Chen, Wangping Ren,

H. Eisaki, S. Uchida, P.J. Hirschfeld and J.C. Séamus Davis

## Supplementary Note 1

## Renormalized mean-field theory of the extended t-J model

The extended *t-J* model on a square lattice is given by

$$H = -\sum_{(i,j),\sigma} P_G t_{ij}\left(c_{i\sigma}^\dagger c_{j\sigma} + h.c.\right)P_G + J\sum_{<i,j>} \boldsymbol{S}_i \boldsymbol{S}_j, \tag{1}$$

where $c_{i\sigma}^\dagger$ creates an electron at the lattice site $i$ with spin $\sigma$. The hopping amplitude $t_{ij}$ is taken to be $t$ and $t'$ when $i, j$ are the nearest-neighbor (NN) and next-nearest-neighbor (NNN) sites, respectively. $< i, j >$ and $(i, j)$ denotes only NN, and both NN and NNN sites, respectively. The Gutzwiller projector $P_G$ projects out all configurations with doubly occupied sites from Hilbert space. Finally, $\boldsymbol{S}_i$ represents the spin operator at site $i$, and $J$ is the superexchange coupling between spins residing at NN sites. The no-double-occupancy constraint can be implemented by employing the Gutzwiller approximation, in which the projection operator $P_G$ is replaced by site-dependent Gutzwiller renormalization factors $g^t$ and $g^s$ for hoppings and superexchange coupling, respectively. The resulting renormalized Hamiltonian now reads,

$$H = -\sum_{(i,j),\sigma} g_{ij}^t t_{ij}\left(c_{i\sigma}^\dagger c_{j\sigma} + h.c.\right) + J\sum_{<i,j>}[g_{ij}^{S,Z} S_i^Z S_j^Z + g_{ij}^{S,xy}(\frac{S_i^+ S_j^- + S_i^- S_j^+}{2})] \tag{2}$$

Further progress can be made by mean-field decoupling of the renormalized Hamiltonian in density and pairing channels with ensuing mean-fields hole density $\delta_i$, bond-field $\chi_{ij\sigma}$, magnetic moment $m_i$, and pair potential $\Delta_{ij\sigma}$ defined as

$$\Delta_{ij\sigma} = \sigma < \Psi_0|c_{i\sigma} c_{j\bar{\sigma}}|\Psi_0 >, \tag{3}$$

$$\chi_{ij\sigma} = < \Psi_0|c_{i\sigma}^\dagger c_{j\sigma}|\Psi_0 >, \tag{4}$$

$$\delta_i = 1 - < \Psi_0|\sum_\sigma n_{i\sigma}|\Psi_0 >, \tag{5}$$

$$m_i = < \Psi_0|S_i^Z|\Psi_0 >, \tag{6}$$

where, $|\Psi_0 >$ is the unprojected ground state wavefunction. A direct diagonalization of the resulting mean-field Hamiltonian will not yield the lowest energy state, however, as the Gutzwiller factor themselves depend on the local mean-fields. Instead, the ground state energy $E_g = <\Psi_0|H|\Psi_0>$ has to be minimized with respect to $|\Psi_0>$ under constraints that the total electron density is fixed and $|\Psi_0>$ is normalized[1]. This leads to following renormalized mean-field Hamiltonian for paramagnetic states ($m_i = 0$).

$$H_{MF} = \sum_{(i,j),\sigma} \epsilon_{ij\sigma} c_{i\sigma}^{\dagger} c_{j\sigma} + h.c. + \sum_{<i,j>,\sigma} \sigma D_{ij\sigma}^{*} c_{i\sigma} c_{j\bar{\sigma}} + h.c. - \sum_{i,\sigma} \mu_{i\sigma} n_{i\sigma}, \qquad (7)$$

where,

$$\epsilon_{ij\sigma} = -g_{ij}^{t} t_{ij} - \delta_{ij,<ij>} \frac{3}{4} J g_{ij}^{S} \chi_{ij\sigma}^{*} \qquad (8)$$

$$D_{ij\sigma} = -\delta_{ij,<ij>} \frac{3}{4} J g_{ij}^{S} \Delta_{ij\sigma} \qquad (9)$$

$$\mu_{i\sigma} = \mu + \frac{3}{4} J \sum_{j\sigma'} (|\Delta_{ij\sigma'}|^2 + |\chi_{ij\sigma'}|^2) \frac{dg_{ij}^{S}}{dn_{i\sigma}} + t_{ij} \sum_{j\sigma'} (\chi_{ij\sigma'} + \chi_{ij\sigma'}^{*}) \frac{dg_{ij}^{t}}{dn_{i\sigma}} \qquad (10)$$

Here, $\delta_{ij,<ij>} = 1$ for NN sites and 0 otherwise. In this work, we have focused only on paramagnetic states since we are interested in charge ordering without any long-range spin ordering as very few experiments suggested the presence of any long-range magnetic order coexisting with charge order in $Bi_2Sr_2CaCu_2O_{8+\delta}$. In this scenario, the Gutzwiller renormalization factors are simply given by the following expressions[1]

$$g_{ij\sigma}^{t} = g_{ij}^{t} = g_i^t g_j^t; g_i^t = \sqrt{\frac{2\delta_i}{1-\delta_i}} \qquad (11)$$

$$g_{ij}^{S,z} = g_{ij}^{S,xy} = g_{ij}^{S} = g_i^S g_j^S; \ g_i^S = \frac{2}{1+\delta_i} \qquad (12)$$

Here, we have assumed that the above expressions are valid at all temperatures of interest[2]. In other words, we have approximated $T \neq 0$ Gutzwiller factors by their values at $T = 0$. Temperature effects enter the calculations via Fermi functions used in the evaluation of the mean-fields [Supplementary Eq. (3-6)]). The renormalized mean-field Hamiltonian in Supplementary Eq. (7) can be diagonalized by using a spin-generalized Bogoliubov transformation, yielding the following Bogoliubov-de Gennes (BdG) equation

$$\sum_j \begin{pmatrix} \epsilon_{ij\uparrow} & D_{ij\uparrow} \\ D_{ji\uparrow}^{*} & -\epsilon_{ij\downarrow} \end{pmatrix} \begin{pmatrix} u_j^n \\ v_j^n \end{pmatrix} = E_n \begin{pmatrix} u_i^n \\ v_i^n \end{pmatrix}. \qquad (13)$$

The BdG equation has to be solved self-consistently as the matrix elements depend on the mean-fields, which, in turn, depend on the eigenvalues ($u_i^n, v_i^n$) and eigenvectors $E_n$. The paramagnetic ground state of the *t-J* model treated within aforementioned renormalized mean-field theory (RMFT) is a uniform d-wave superconductor (DSC). However, we are interested in pair density wave (PDW) solutions which have been shown to be very close in energy to the DSC state within RMFT[1-3] as well as in more rigorous numerical schemes like variational Monte-Carlo[4,5] and tensor networks[6]. PDW states can be obtained by initializing BdG equation [Supplementary Eq. (13)] with modulating pair-field (keeping other mean-fields uniform) with the following form.

$$\Delta_{i,i+\hat{x}} = \Delta_0 + \Delta_Q \cos(\boldsymbol{Q}_P \cdot \mathbf{R}_i) \tag{14}$$

$$\Delta_{i,i+\hat{y}} = -\Delta_0 - \Delta_Q \cos(\boldsymbol{Q}_P \cdot \mathbf{R}_i - \frac{\hat{x}}{2}) \tag{15}$$

Here, the modulation wavevector is chosen to be $\boldsymbol{Q}_P = (\pm 1/8, 0)\frac{2\pi}{a_0}$ based on experimental evidences[7,8,9]. A bond-centered PDW state with coexisting d-wave superconductivity (PDW+DSC) can be obtained as a self-consistent solution using a finite $\Delta_0 < \Delta_Q$, whereas a pure PDW state can be obtained by setting $\Delta_0 = 0$ in the initial seed. We note that a computationally more efficient scheme to study unidirectional modulating states (in absence of disorder) is obtained by exploiting translational invariance in direction orthogonal to modulations. Here, BdG equations on 2D lattice are Fourier transformed in the orthogonal direction to yield quasi-1D BdG equations. Details of this scheme can be found in Ref. [3]. Fig. 2(a), (c)-(f) in the main-text have been obtained using this scheme.

Results presented in the main-text were obtained using the parameter set $t = 400$ meV, $t' = -0.3t$, $J = 0.3t$. Further, we chose hole-doping $p = 0.125$, which is larger than the doping $p = 0.08$ at which experiments discussed in the main-text were performed because of the following reasons. First, it has long been known that the *t-J* model overestimates the doping scale of DSC dome by almost a factor of two. Experiments find the DSC dome to be in hole doping range $p \sim 0.05$-$0.3$ whereas in the RMFT *t-J* model (with the aforementioned parameter set) it turns out to be in the range $0.01$-$0.45$ (at $T = 0$)[3]. If we account for this scale difference, then $p = 0.125$ will be closer to the experimental doping of $p = 0.08$. Second, it's hard to get converged PDW solutions at very low dopings as the derivatives of Gutzwiller factors, entering in the on-site potentials [Supplementary Eq. (10)], fluctuate strongly even with a small change in local doping[10]. This is more severe when solving the impurity problem. Finally, our conclusions mainly depend on just one premise: the low-temperature state is PDW+DSC and the high-temperature state is pure PDW, which does not depend on the actual doping level as long as it remains below a critical level ($p \sim 0.18$ at $T = 0$) to realize these states.

For the aforementioned parameter set, a self-consistent pure PDW state is obtained in temperature range $0 < T < 0.11t$ whereas the PDW+DSC state is found as a stable solution for $0 < T < 0.085t$. Both PDW and PDW+DSC states have almost equal energy per site, which is a few meV larger than the uniform DSC state[1,2,3,11]. This tiny energy difference between PDW+DSC and DSC state can be overcome by a variety of means, such as disorder which is not accounted for in the calculation. We have effectively assumed such effects to be present, leading to the PDW+DSC state at low-temperatures (in the range $0 < T < 0.085t$) and the pure PDW state at higher temperatures (in the range $0.085t < T < 0.11t$). In the PDW+DSC state, increasing temperature leads to a sharp decrease in the uniform DSC component ($\Delta(\boldsymbol{q}=\boldsymbol{0})$) as shown in the main-text Figure 2e. For $0.05t < T < 0.085t$, $\Delta(\boldsymbol{q}=\boldsymbol{0})$ is very small but finite. This 'fragile PDW+DSC' state is a stable solution of the RMFT equations and not a computational artefact. This result is verified by the observation that lowering the self-consistency tolerance by an order of magnitude yields the same state.

In the main-text Fig. 2c and 2d, we showed spatial variation of hole density and d-wave gap order parameter, respectively. To complete the discussion of mean-fields, Supplementary Figure 2a-c show the spatial variation of NN bond mean-field $\chi_{ij}$ in PDW+DSC (at $T = 0.01t, 0.04t$) and pure PDW state (at $T = 0.09t$). We find that the modulations in $\chi_{ij}$ are typically of the size ~0.1t in PDW+DSC at low temperatures and ~0.05t in pure PDW states at higher temperatures. The bare bond fields are not physical observables, however. The physical expectation value of the bond operator ($c_{i\sigma}^{\dagger} c_{j\sigma}$) in the Gutzwiller projected state is the bond mean-field scaled by the Gutzwiller hopping factor: $\chi_{ij}^{o} = g_{ij}^{t}\chi_{ij}$ [12]. We can define the NN bond order at a given lattice site $i$ as $\chi_{i}^{o} = (\chi_{i,i+\hat{x}}^{o} + \chi_{i,i-\hat{x}}^{o} + \chi_{i,i+\hat{y}}^{o} + \chi_{i,i-\hat{y}}^{o})/4$, where $i \pm \hat{x} (\hat{y})$ represent NN sites to $i$ along x(y)-direction. As evident from Supplementary Figure 2d, the size of modulations in the bond order turns out to be an order of magnitude smaller than the bare mean-field. Similar to the case of hole density, the reduction in the modulation amplitude of bond variables for higher temperatures is a consequence of the reduction in the PDW gap order parameter (Fig. 2e in the main-text). Finally, we note that the bond order in both PDW and PDW+DSC states has a dominant $d$-form factor[11].

In order to compute local density of states (LDOS), we first obtain lattice Green's functions $G_{ij}(E)$ using the eigenvalues and eigenvectors of the BdG matrix [Supplementary Eq. (13)].

$$G_{ij}(E) = g_{ij}^{t} \sum_{n} \frac{u_{i}^{n} u_{j}^{n*}}{\omega - E_{n} + i0^{+}} \tag{16}$$

Here, $0^{+}$ is a small artificial broadening set to be $0.01t$, and the sum runs over all the eigenvalues. The diagonal lattice Green's function yields total LDOS at a site:

$$N_{i}(E) = -\frac{2}{\pi} Im[G_{ii}(E)], \tag{17}$$

where, $Im$ represents imaginary part and the factor 2 accounts for spin degeneracy. Differential conductance measured in an STM experiment is, however, proportional to the sample's LDOS evaluated at the STM tip position[13]. Thus, we must compute the continuum LDOS few angstroms above the exposed BiO layer in $Bi_2Sr_2CaCu_2O_{8+\delta}$ for a meaningful comparison with the experimental data. Accordingly, we obtain continuum Green's function $G(\boldsymbol{r}, \boldsymbol{r}'; E)$ via a basis transformation[14] from lattice to continuum space where the matrix elements of the transformation are given by the Wannier function $W_i(\boldsymbol{r})$ centered at lattice site $i$.

$$G(\boldsymbol{r}, \boldsymbol{r}'; E) = \sum_{ij} W_i(\boldsymbol{r}) G_{ij}(E) W_j^*(\boldsymbol{r}') \tag{18}$$

The imaginary part of the diagonal continuum Green's function yields LDOS at a continuum point $\boldsymbol{r}$.

$$N(\boldsymbol{r}, E) = -\frac{2}{\pi} Im[G(\boldsymbol{r}, \boldsymbol{r}; E)] \tag{19}$$

We have obtained the continuum LDOS at a height $\sim 4\text{Å}$ above the BiO layer in $Bi_2Sr_2CaCu_2O_{8+\delta}$ employing a first-principles Cu-3$d_{x2-y2}$ Wannier function obtained using the Wannier90 package, identical to that used in Ref. [3,15] and very similar to that in Ref. [16].

Supplementary Fig. 1a shows the continuum LDOS map at $E = \Delta_1$ in the pure PDW state at $T = 0.09t$. The LDOS shows a periodicity of $4a_0$. Supplementary Fig. 1b shows spectra at eight Cu positions marked in the panel 1a. Sharp features present at higher energies are expected to be broadened by inelastic scattering, which has been shown in Ref. [17] to be essential to account for the spectral lineshapes in underdoped cuprates. In that work, it was shown that the effects of inelastic scattering can be simply incorporated by adding a linear-in-energy term $i\Gamma = i\alpha|E|$ to the constant artificial broadening $i0^+$ used in calculation of lattice Green's function [Supplementary Eq.(16)]. Using the experimental fits presented in Ref. [17], we set $\alpha = 0.25$. Supplementary Fig. 1c shows the continuum LDOS incorporating the linear inelastic scattering. All LDOS, $Z(\boldsymbol{q}, E)$, and $\Lambda_P(\boldsymbol{q}, \Delta_0)$ results presented in the main-text, and gap map results presented in Supplementary Figure 7 have been obtained after accounting for the inelastic scattering.

We note that a finite value of artificial broadening $i0^+$ used in our calculations is responsible for non-zero LDOS at zero bias in the PDW+DSC state, as seen in Fig. 2a. Indeed, with decreasing artificial broadening, the zero-bias LDOS in PDW+DSC state approaches 0 due to presence of nodes in the quasiparticle spectrum[18], as evident from Supplementary Figure 3. On the contrary, the zero-bias LDOS saturates at a finite value in pure PDW state due to the presence of Bogoliubov-Fermi surface[1,18].

## Supplementary Note 2

## PG gap $\Delta_1(r)$ modulation detection

We determine the gap map $\Delta_1(r)$ by measuring the energy of the coherence peak in each $dI/dV$ spectrum at $E > 0$. Supplementary Figure 5b shows the magnitude of the power-spectral-density Fourier transform $\Delta_1(\boldsymbol{q})$ of the gap map $\Delta_1(r)$ in Figure 5a. There is strong disorder in $\Delta_1(\boldsymbol{q})$ surrounding $\boldsymbol{q} = 0$. The feature at a length of 1/5 in the (0, 0)-(1, 1) direction is related to the BiO supermodulation. The feature at about 20 degrees off the (0, 0)-(1, 0) direction at a length about 1/6 is possibly related to the electronic disorder. In Supplementary Figure 5, we show $\Delta_1(\boldsymbol{q})$ intensities before and after the exponential background has been subtracted. After the background is subtracted, the maxima at $\boldsymbol{Q}_P \approx (0, \pm1/8)2\pi/a_0$ and $\boldsymbol{Q}_P \approx (\pm1/8, 0)2\pi/a_0$ become clearly visible. This analysis provides one type of experimental evidence of the $8a_0$ modulations in $\Delta_1(\boldsymbol{r})$.

We apply a computationally two-dimensional lock-in technique to obtain the amplitude $\Delta_{q_i}(\boldsymbol{r})$ of the gap modulation $\Delta_1(\boldsymbol{r})$ at $\boldsymbol{q}_i$. $\Delta_1(\boldsymbol{r})$ is multiplied by $e^{i\boldsymbol{q}_i \cdot \boldsymbol{r}}$ and integrated over a Gaussian filter to obtain the complex-values lock-in signal[9,19]

$$\Delta_{q_i}(\boldsymbol{r}) = \frac{1}{\sqrt{2\pi}\sigma} \int d\boldsymbol{R} \Delta_1(\boldsymbol{R}) e^{i\boldsymbol{q}_i \cdot \boldsymbol{R}} e^{\frac{|\boldsymbol{r}-\boldsymbol{R}|^2}{2\sigma^2}} \tag{20}$$

Where $\boldsymbol{q}$ denotes the wavevector of interest and $\sigma$ the average length-scale in $\boldsymbol{r}$-space. This technique is implemented in $\boldsymbol{q}$-space

$$\Delta_{q_i}(\boldsymbol{r}) = \mathcal{F}^{-1} \Delta_{q_i}(\boldsymbol{q}) = \mathcal{F}^{-1} [\mathcal{F}(\Delta_1(\boldsymbol{r}) e^{i\boldsymbol{q}_i \cdot \boldsymbol{r}}) \cdot \frac{1}{\sqrt{2\pi}\sigma_q} e^{-\frac{q^2}{2\sigma_q q^2}}] \tag{21}$$

where $\sigma_q = 1/\sigma$ is the cut-off length in $\boldsymbol{q}$-space. $\sigma$ is specified to capture only the relevant image distortions.

**Supplementary Note 3**

**Atomic precision image registration**

In the temperature dependence experiments, $T(\boldsymbol{r}, 5\text{ K})$ and $T(\boldsymbol{r}, 55\text{ K})$ are measured in the same field of view with sub-unit-cell resolution. The data are processed by performing the Lawler-Fujita procedure[20] that maps the data onto a perfectly periodic lattice without lattice distortions. The data are subsequently corrected using shear transformation to maintain the C4 symmetry of the $CuO_2$ crystal lattice. After the topographs are corrected, $T(\boldsymbol{r}, 5\text{ K})$ and $T(\boldsymbol{r}, 55\text{ K})$ are registered to the exact same FOV with atom-by-atom precision as shown in Supplementary Figure 6a and b. Subtraction of $T(\boldsymbol{r}, 5\text{ K})$ from $T(\boldsymbol{r}, 55\text{ K})$ gives rise to $\delta T(\boldsymbol{r})$ in Supplementary Figure 6c. The differences between $T(\boldsymbol{r}, 5\text{ K})$ and $T(\boldsymbol{r}, 55\text{ K})$ are noise and distortions in individual unit cells. They are not relevant to the demonstration from $\delta T(\boldsymbol{r})$ that the FOVs of 5 K and 55 K are identical.

The differential conductance map $g(\boldsymbol{r}, V)$ is simultaneously acquired with $T(\boldsymbol{r})$. Applying the same image processing procedures of correcting $T(\boldsymbol{r})$ to $g(\boldsymbol{r}, V)$ gives rise to the temperature induced electronic structure changes. The electronic structures are measured in a wide energy range from -800 mV to 800 mV which includes the PG energy range. The cross-correlation coefficient between $g(\boldsymbol{r}, V)$ at 5 K and 55 K are around 0.9 in the large energy range (Supplementary Figure 6d). This method provides meaningful subtraction of high ($T > T_c$) and low ($T < T_c$) temperature data to detect temperature induced differences of the electronic structures at atomic scale.

**Supplementary Note 4**

**Predicted temperature-evolution of gap map $\Delta_1(\boldsymbol{r})$**

We calculated the temperature evolution of the gap map $\Delta_1(\boldsymbol{r})$. $\Delta_1(\boldsymbol{r})$ is defined as the energy of the coherence peak at $E > 0$, i.e., the same definition as the experimental measurement in main-text Figure 3. The gap modulation in the PDW+DSC state has a periodicity of $8a_0$ (Supplementary Figure 7a and b). The amplitude of the y-averaged gap modulation is $\sim 0.14t$ at $T = 0$ and $\sim 0.13t$ at $T = 0.04t$. The gap modulation in the pure PDW state has a periodicity of $4a_0$ (Supplementary Figure 7c). The amplitude of the y-averaged gap modulation is $\sim 0.05t$ at $T = 0.09t$, which is much smaller compared to the PDW+DSC state. This is a consequence of the reduction in the PDW gap order parameter with increasing temperature (Fig. 2d). In this prediction the modulation periodicity of $\Delta_1(\boldsymbol{r})$ changes from $8a_0$ to $4a_0$ in the transition from the PDW+DSC state to the pure PDW state. In experiments we have observed that $\Delta_1(\boldsymbol{r})$ modulates at $8a_0$ (inset in main-text Figure 3a) at $T \ll T_c$. However, the modulation periodicity of $\Delta_1(\boldsymbol{r})$ could not be determined at $T = 55\text{K} = 1.5T_c$ due to the presence of large regions with indeterminate coherence peaks (see white regions in Figure 3f). Therefore, the predicted temperature-evolution of the gap map $\Delta_1(\boldsymbol{r})$ could not be tested.

We note that the gap modulation is also possible in a state with coexisting charge density wave (CDW) and uniform DSC. In particular, a $d$-form factor (dFF) bond density wave (BDW) with wavevector $\boldsymbol{Q}_c = (\pm 1/4, 0)\frac{2\pi}{a_0}$ is often considered as a main candidate of charge order in underdoped cuprates both below and above $T_c$[21,22,23]. These states (Supplementary Figure 8), however, cannot account for the presence of gap modulations with wavevectors $\sim (\pm 1/8, 0)\frac{2\pi}{a_0}$ and $(0, \pm 1/8)\frac{2\pi}{a_0}$ in the experimental data (main-text Fig. 3a).

**Supplementary Note 5**

**Bogoliubov quasiparticle scattering interference calculations**

Bogoliubov quasiparticle scattering interference (BQPI) is a consequence of impurity scattering. To study the BQPI characteristics of the PDW+DSC and PDW states, we consider a point-like potential scatterer with impurity potential $V_{imp}$ located at the lattice site $i^*$ in the middle of an $N \times N$ square lattice. The resulting system is described by the following Hamiltonian

$$H = H_{MF} + H_{imp}, \tag{22}$$

where, $H_{MF}$ is given by Supplementary Eq. (7), and the impurity Hamiltonian can be expressed as

$$H_{imp} = V_{imp} \sum_\sigma n_{i^*\sigma}, \tag{23}$$

We set $N = 56$ and $V_{imp} = 3t$. The Hamiltonian $H$ can be diagonalized following the same procedure used for diagonalizing $H_{MF}$. The resulting BdG equations have the same form as the clean system [Supplementary Eq. (13)] with only difference that the onsite potentials [Supplementary Eq. (10)] are changed to $\mu_{i\sigma} \rightarrow \mu_{i\sigma} - V_{imp}\delta_{ii^*}$, where $\delta_{ij}$ is the Kronecker delta function. We solve the BdG equations self-consistently to obtain the PDW+DSC and pure PDW states in presence of an impurity. Subsequently, we compute continuum LDOS $N(\boldsymbol{r}, E)$ and thereby $Z(\boldsymbol{r}, E) = N(\boldsymbol{r}, +E)/N(\boldsymbol{r}, -E)$ for $E > 0$ using the procedure outlined in Supplementary Note 1. QPI $Z(\boldsymbol{q}, E)$ maps are obtained by taking Fourier transform of the $Z(\boldsymbol{r}, E)$ maps. Finally, energy integrated BQPI maps are obtained by summing $Z(\boldsymbol{q}, E)$ maps over the range $0 < E < \Delta_0$.

$$\Lambda_P(\boldsymbol{q}, \Delta_0) = \sum_{E \cong 0}^{\Delta_0} Z(\boldsymbol{q}, E) \tag{24}$$

The upper cut-off of the energy sum is set to $\Delta_0 = 0.05t = 20$ meV ($t = 400$ meV) to match with the experiment.

The as obtained $Z(\boldsymbol{q}, E)$ maps exhibit largest intensity at PDW driven charge order Bragg peaks $\boldsymbol{q} = \pm n\boldsymbol{Q}_P, n = 0, 1, 2, \dots, 7$ in PDW+DSC state and $\boldsymbol{q} = \pm n(2\boldsymbol{Q}_P), n = 0, 1, 2, 3$, in the pure PDW state (the fundamental charge order harmonic occurs at $\boldsymbol{Q}_c = \boldsymbol{Q}_P$ in PDW+DSC state and at $\boldsymbol{Q}_c = 2\boldsymbol{Q}_P$ in pure PDW state as explained in the main-text), see Supplementary Fig. 9a, d. Accounting for the discommensurate short-range nature of the charge order in Bi$_2$Sr$_2$CaCu$_2$O$_{8+\delta}$ [24] will smear the Bragg peaks and reduce their intensity. The exact amount of suppression is not clear, though. In order to emphasize the QPI wavevectors emerging from impurity scattering, we have chosen to suppress the Bragg peaks by a factor $F = 100$ (Supplementary Fig. 9b, e). Finally, the resulting $Z(\boldsymbol{q}, E)$ maps are symmetrized by adding their 90°-rotated versions to account for the orthogonal domains of unidirectional charge modulations seen in the experiments [25] (Supplementary Fig. 9c, f). To further illustrate the effects of suppression of charge order Bragg peaks we show $\Lambda_P(\boldsymbol{q}, \Delta_0)$-maps with suppression factors $F = 1, 10, 50, 100, 1000$, in Supplementary Fig. 10. Clearly, if shown to scale ($F = 1$), the Bragg peaks will obscure all wavevectors emerging from impurity scattering. We found that a better match with the experimental result can be obtained by using $F = 100$, although the qualitative features do not change significantly with $F$ once the Bragg peaks are suppressed somewhat, around $F = 50$.

Energy integrated BQPI map $\Lambda_C(\boldsymbol{q}, \Delta_0)$ in CDW state is constructed non-self-consistently via two independent methods. The first method is setting gap order parameter in self-consistent PDW state (at $T = 0.09t$) to zero while keeping bond order and on-site potential modulations intact (Supplementary Fig. 11a and main-text Fig. 5b). The other method is taking the normal state Hamiltonian from the uniform DSC state solution (at $T = 0.09t$) and adding a term producing a $d$-form factor bond ordered charge density wave ($d$FF-BDW) with wavevector $\boldsymbol{Q}_C = (1/4,0)\frac{2\pi}{a_0}$ (Supplementary Fig. 11b). The amplitude of the charge density wave is set to be the same as the uniform DSC state gap field. This state becomes equivalent to that in Supplementary Fig. 8 if the coexisting DSC state in the later is removed. To calculate $\Lambda_C(\boldsymbol{q}, \Delta_0)$, an impurity Hamiltonian is added and subsequently the corresponding total Hamiltonians are diagonalized in the real space. This procedure is equivalent to a T-matrix calculation. $\Lambda_C(\boldsymbol{q}, \Delta_0)$ in the pure CDW state obtained from both methods exhibit features very different from the $\Lambda(\boldsymbol{q}, \Delta_0)$ observed in experiments. We have presented $\Lambda_C(\boldsymbol{q}, \Delta_0)$ map from Supplementary Figure 11a in the main-text Figure 5b.

## Supplementary Note 6

### Comparison between theoretical and experimental $\Lambda(\boldsymbol{q}, \Delta_0)$ data

Here we compare the theoretical and experimental $\Lambda(\boldsymbol{q}, \Delta_0)$ data in detail. The feature extending in the nodal directions (red arrow in Supplementary Figure 12a and c) is a signature of the uniform DSC component in the PDW+DSC state (below $T_c$), which disappear in the pure PDW state (above $T_c$). Supplementary Figure 12e-f shows superimposing the predicted $\Lambda_P(\boldsymbol{q}, \Delta_0)$ of PDW+DSC state onto the measured $\Lambda(\boldsymbol{q}, \Delta_0)$ of the superconducting phase, and superimposing the $\Lambda_P(\boldsymbol{q}, \Delta_0)$ of pure PDW state onto the $\Lambda(\boldsymbol{q}, \Delta_0)$ of the pseudogap phase, respectively. The positions of the experimental and theoretical QPI features are nearly identical.

Moreover, we measure the arc-like feature in the experimental $\Lambda(\boldsymbol{q}, \Delta_0)$ and theoretical $\Lambda_P(\boldsymbol{q}, \Delta_0)$. The extension of the arc is quantified by the angle subtended by the arc. We fit each arc of a circle about $(\pm 1, \pm 1)\ 2\pi/a_0$ point using least square fit (see Supplementary Figure 13). This procedure is carried out for six temperatures in both the theory and the experiment. The measured arc extension increases as a function of temperature from superconducting to pseudogap phase (Supplementary Figure 14a). This measurement agrees with the predicted arc extension in $\Lambda_P(\boldsymbol{q}, \Delta_0)$ from PDW+DSC state to pure PDW state (Supplementary Figure 14b).

## Supplementary Note 7

### Energy-evolution of QPI signatures of the pseudogap phase and the PDW state

To avoid the 'setup' effect in the experiments, we calculate the ratio of the total density of states

$$Z(\boldsymbol{r}, \text{V}) \equiv \frac{g(\boldsymbol{r}, +\text{V})}{g(\boldsymbol{r}, -\text{V})} \tag{25}$$

We take the power spectral density Fourier transform $Z(\boldsymbol{q}, \mathrm{V})$ of $Z(\boldsymbol{r}, \mathrm{V})$. The $Z(\boldsymbol{q}, \mathrm{V})$ are summed up to $\Delta_0$, the energy that the Bogoliubov quasiparticles cease to exist[26]. $\Delta_0$ is around 20 meV in the 8% hole-doped $Bi_2Sr_2CaDyCu_2O_8$ sample studied in this paper. The energy evolution of the experimental $Z(\boldsymbol{q}, \mathrm{V})$ maps from 8 meV to 20 meV and the corresponding calculated $Z(\boldsymbol{q}, \mathrm{V})$ maps are presented in Supplementary Figure 15. The energy evolution of the wavevectors are visualized in a supplementary movie of $Z(\boldsymbol{q}, V, 55 \text{ K})$ from 2 mV to 20 mV. The wavevectors evolve dispersively with energy only by a small amount.

**Legends of Additional Supplementary Files**

Supplementary Movie 1. Determination of $\Delta_0$ from a movie of $Z(\boldsymbol{q}, V)$ at $T = 4.2$ K. $\Delta_0$ is defined as the energy that the Bogoliubov quasiparticles cease to exist.

Supplementary Movie 2. Energy evolution of quasiparticles in the pseudogap phase shown in a movie of of $Z(\boldsymbol{q}, V)$ at $T = 55$ K.

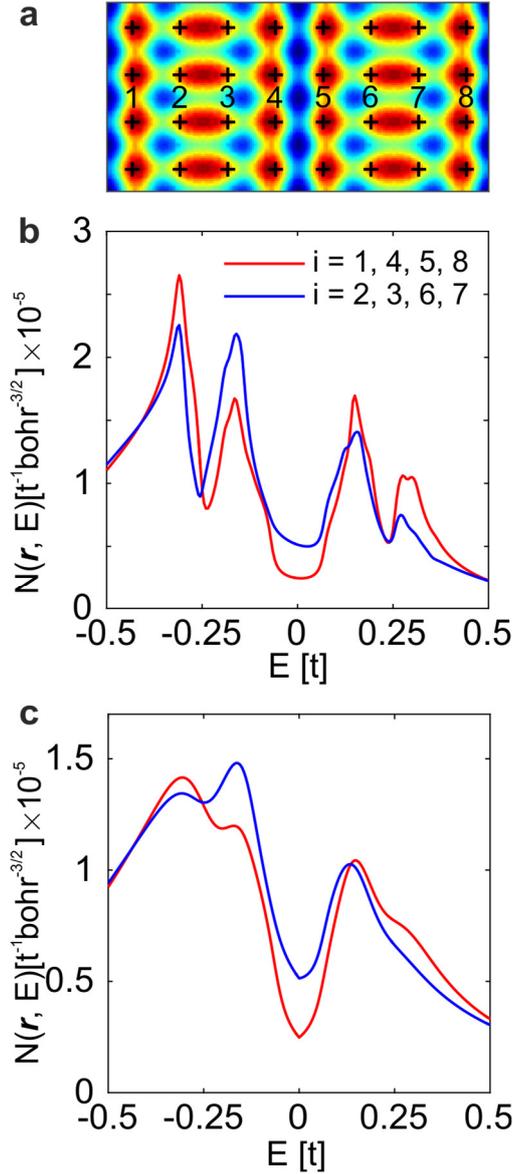

**Supplementary Figure 1. Continuum LDOS in PDW state over a period of PDW (8a$_0$).**
**a.** Continuum LDOS map $N(\boldsymbol{r}, E = \Delta_1)$ in the pure PDW state at $T = 0.09t$ over a 4a$_0$×8a$_0$ area, obtained using Supplementary Eq. (19), for the same parameter set as in Fig. 2a of the main-text. The location of a Cu atom is indicated by a black cross.
**b.** Continuum LDOS spectra $N(\boldsymbol{r}, E)$ above Cu positions, marked in the panel (a), without incorporating inelastic scattering.
**c.** Continuum LDOS spectra $N(\boldsymbol{r}, E)$ at the same positions as in (b), obtained after incorporating $\Gamma = \alpha|E|$ inelastic scattering ($\alpha = 0.25$).

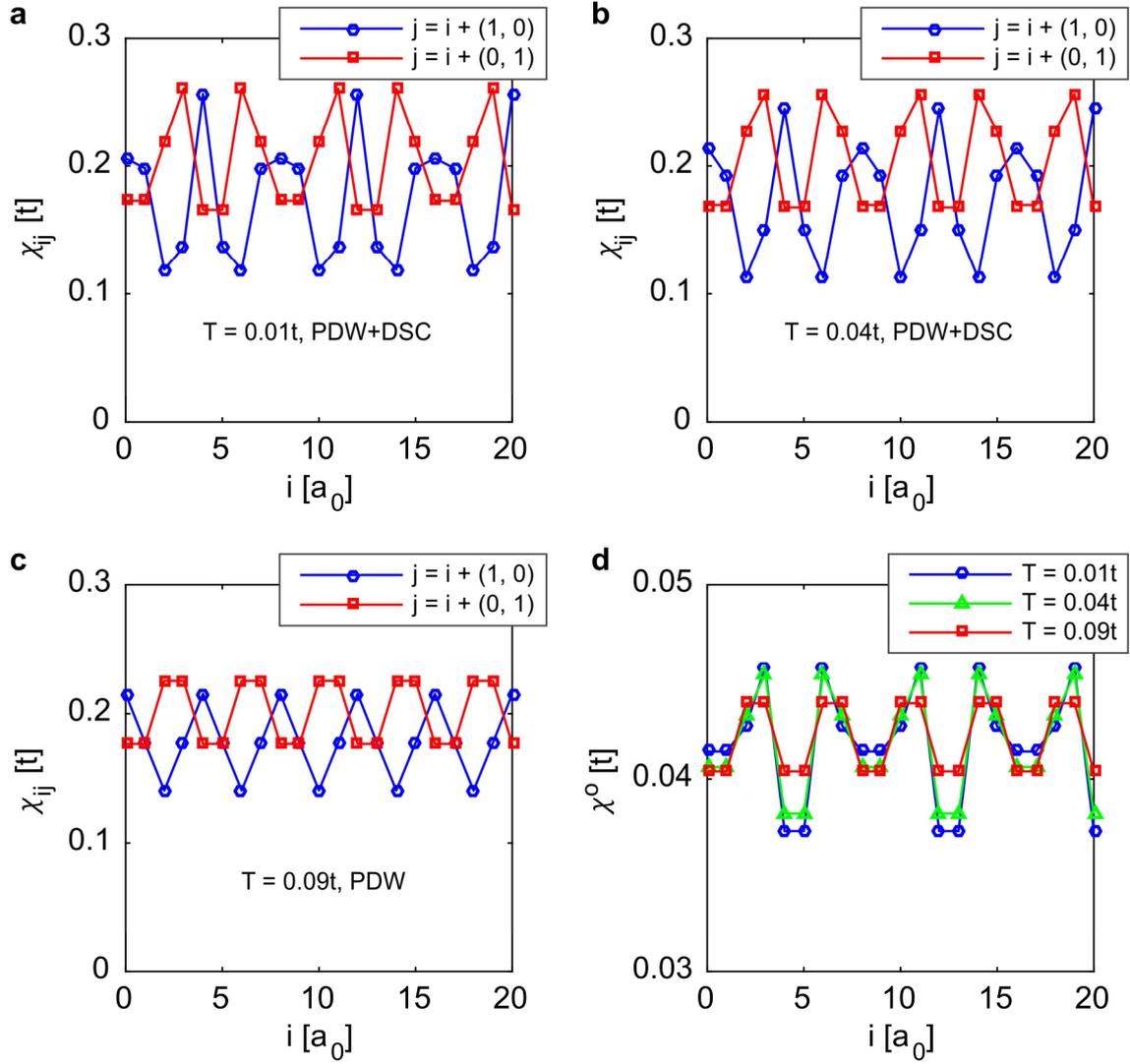

**Supplementary Figure 2. Variation of bond mean-fields and bond order in PDW+DSC and pure PDW states.** Nearest-neighbor bond mean-fields in PDW+DSC state at (**a**) $T = 0.01t$, (**b**) $T = 0.04t$, and (**c**) in pure PDW state at $T = 0.09t$; and (**d**) bond order parameter $\chi_i^o$ at these temperatures.

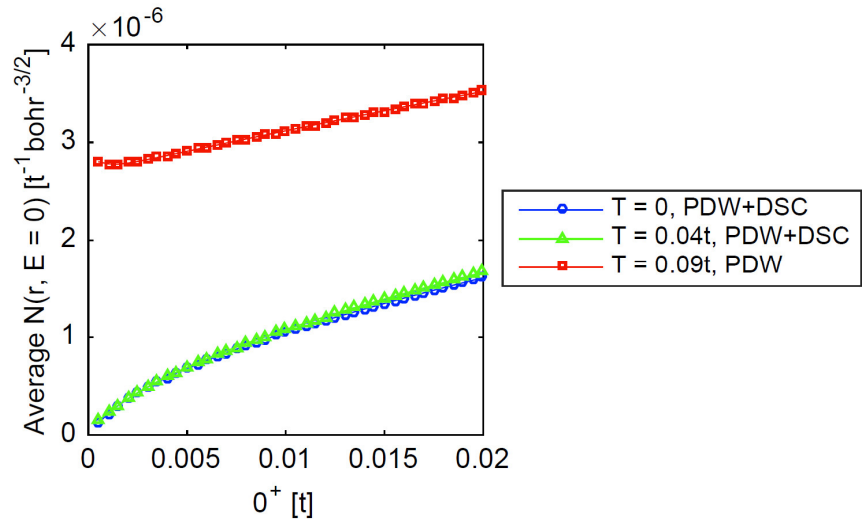

**Supplementary Figure 3**. Variation of average continuum LDOS at zero-energy with artificial broadening factor $0^+$ employed in the calculation of the lattice Green's function [Supplementary Eq. (16)].

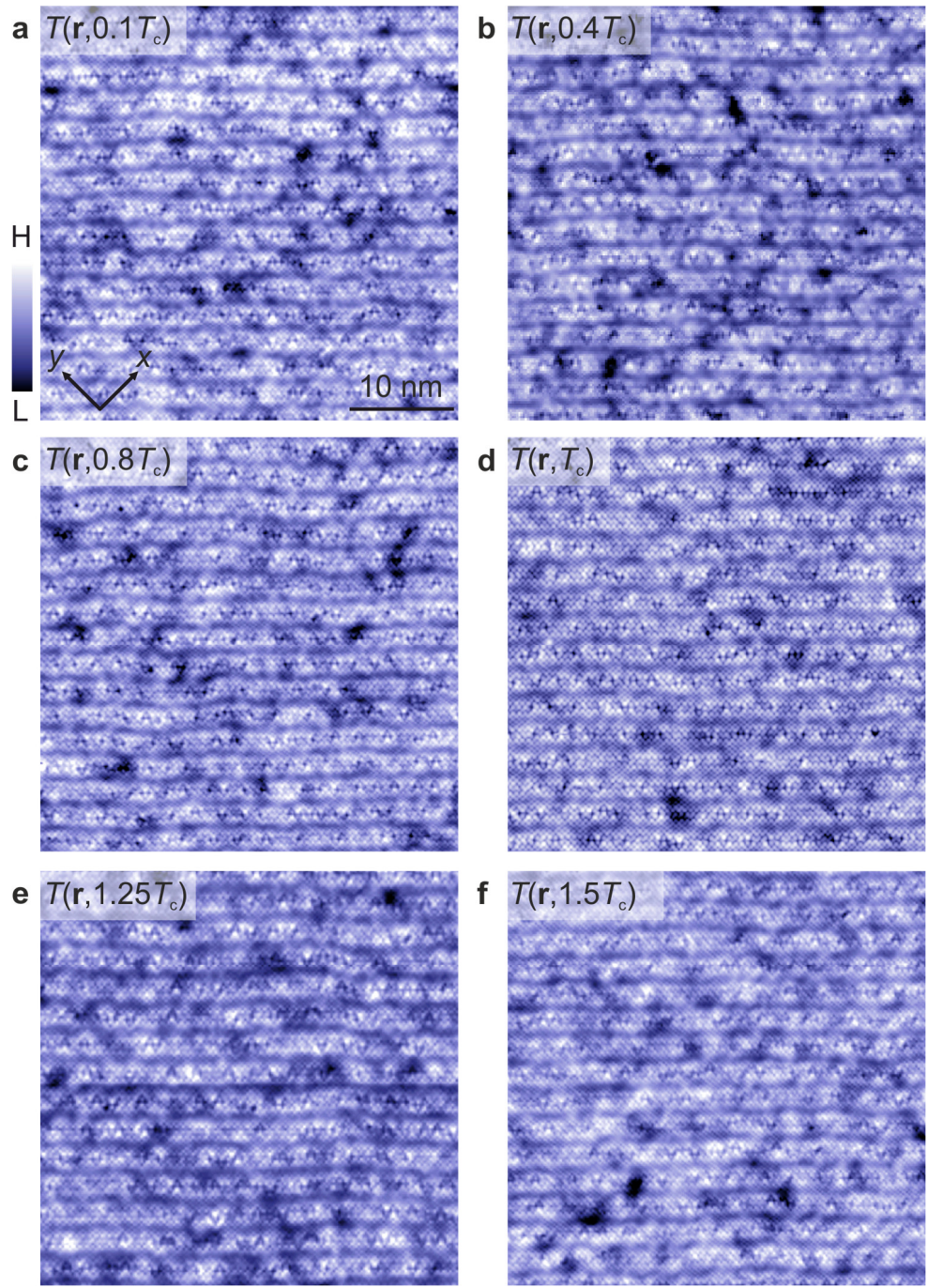

**Supplementary Figure 4**. Topography $T(\mathbf{r})$ in a 40 nm × 40 nm FOV of the underdoped Bi₂Sr₂CaDyCu₂O₈ sample. The six QPI $\Lambda(\boldsymbol{q}, \Delta_0)$ maps from $0.1T_c$ to $1.5T_c$ are measured therein.

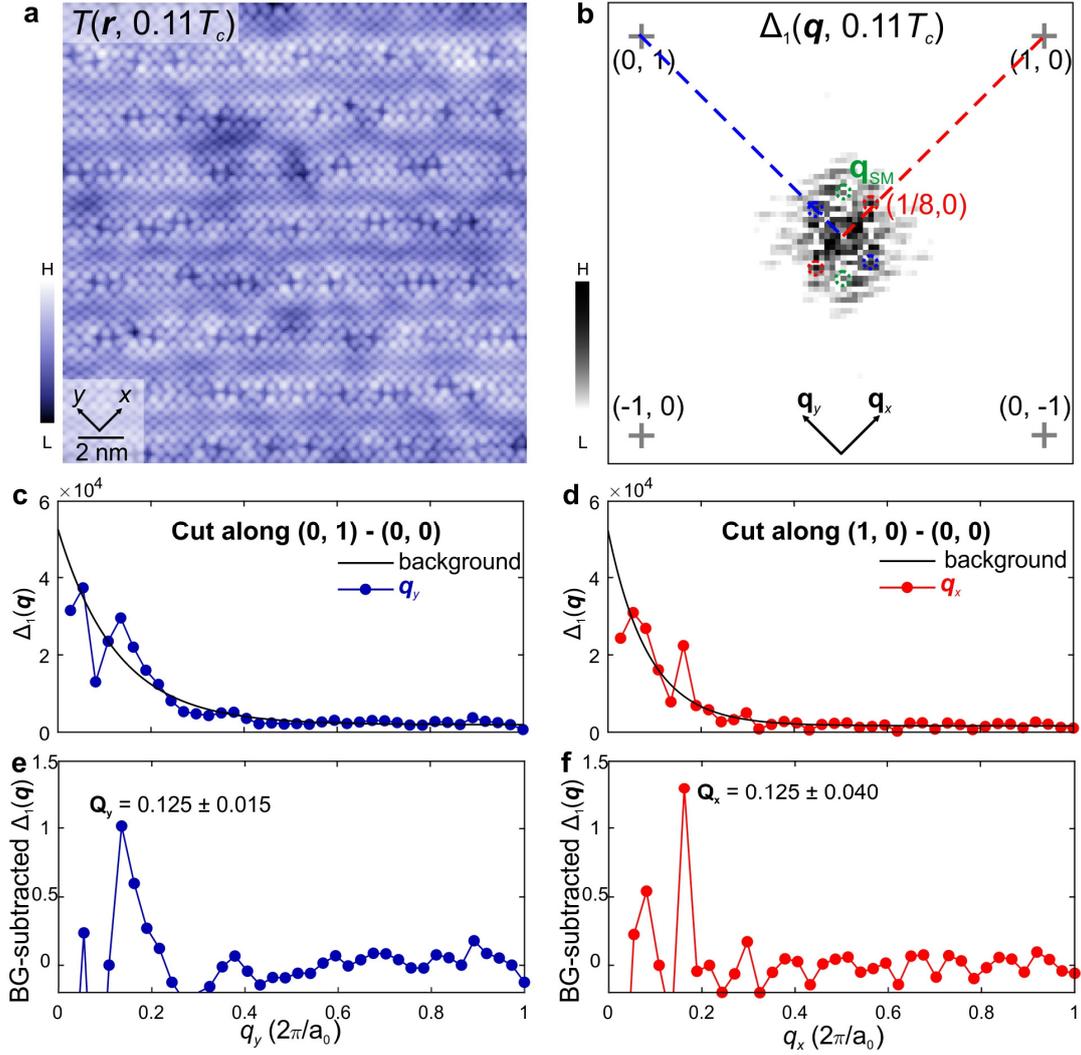

**Supplementary Figure 5. Spatial variations and modulations in cuprate pseudogap energy gaps.**

**a**. 20 nm × 20 nm topographic image $T(\boldsymbol{r})$ of BiO termination of $p \approx 8\%$ hole-doped $Bi_2Sr_2CaDyCu_2O_8$ surface at $T = 0.11T_c = 4.2$K. The gap map $\Delta_1(\boldsymbol{r})$ in Figure 3a in the main-text is taken simultaneously from this FOV.

**b**. Amplitude Fourier transform $\Delta_1(\boldsymbol{q})$ derived from the gap map $\Delta_1(\boldsymbol{r})$ at $T = 0.11T_c = 4.2$K (Figure 3a in the main-text). The 1/8 peaks are marked by blue and red circles. The $\sim 1/5$ peaks related to the supermodulation of the BiO termination are marked by green circles.

**c** & **d**. Measured $\Delta_1(\boldsymbol{q})$ along (**c**) (0,0)-(0,1) and (**d**) (0,0)-(1,0). The linecut measurements are transverse average of 2 or 3 pixels. The measurements are subsequently fitted to an exponential background.

**e** & **f**. The same data as **c** & **d** but with the exponential background subtracted. The intensity due to the PG gap modulation is strongest at the $\boldsymbol{Q}_p \approx (0, \pm1/8)2\pi/a_0$ that represents eight-unit-cell gap modulations in the $y$ direction, and $\boldsymbol{Q}_p \approx (\pm1/8, 0)2\pi/a_0$ that represents the gap modulations in the $x$-direction.

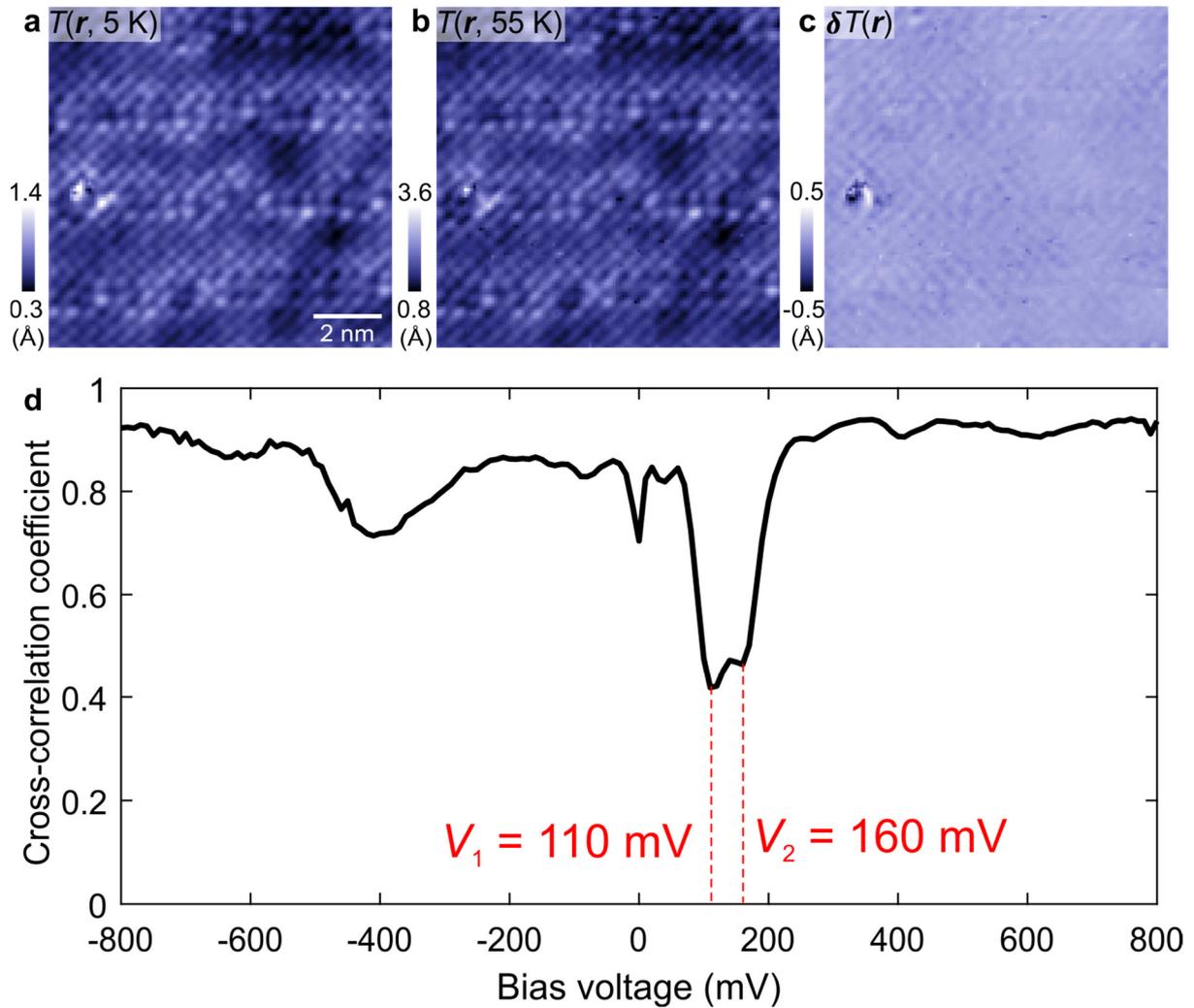

**Supplementary Figure 6. Spatial registration of the two datasets at 5 K and 55 K.**

**a** & **b**. $T(\mathbf{r}, 5\,\text{K})$ and $T(\mathbf{r}, 55\,\text{K})$ processed using the Lawler-Fujita algorithm. The distortions are now corrected and the $CuO_2$ lattice are identically periodic.

**c**. Measured $\delta T(\mathbf{r}) = T(\mathbf{r}, 55\,\text{K}) - T(\mathbf{r}, 5\,\text{K})$ showing the FOVs are identical.

**d**. Cross-correlation coefficient between $g(\mathbf{r}, V, 5\,\text{K})$ and $g(\mathbf{r}, V, 55\,\text{K})$ as a function of bias voltage. There is strong correspondence between the two $g(\mathbf{r}, V)$ maps in the large energy scale except the PG energy gap scale from 110 to 160 mV.

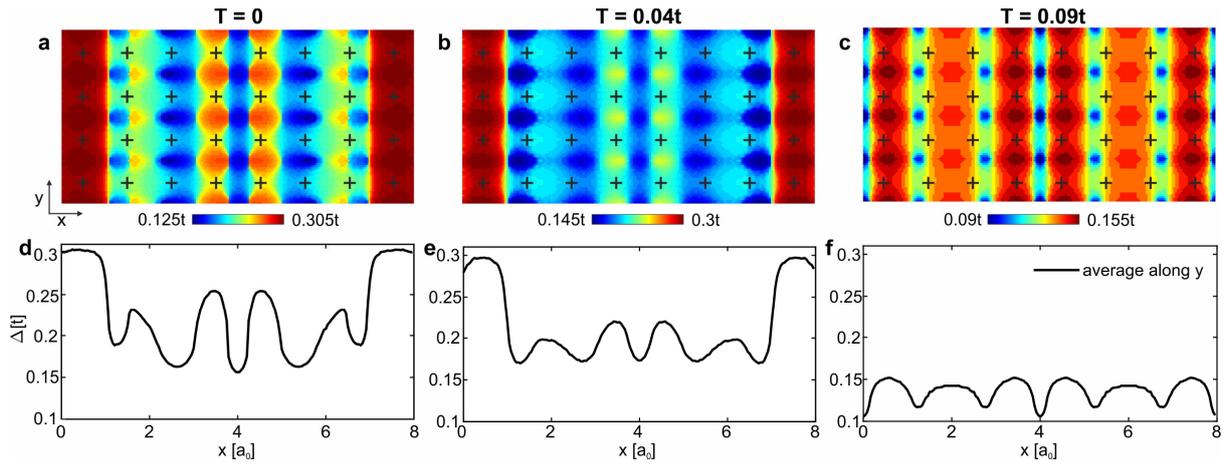

**Supplementary Figure 7. Predicted temperature evolution of $\Delta_1(r)$ gap maps.**
**a-c**. Gap maps in PDW+DSC state at (**a**) $T = 0$, (**b**) $T = 0.04t$, and in pure PDW state at (**c**) $T = 0.09t$ over $8a_0 \times 4a_0$ area.
**d-f**. Gap averaged along y-axis (black).

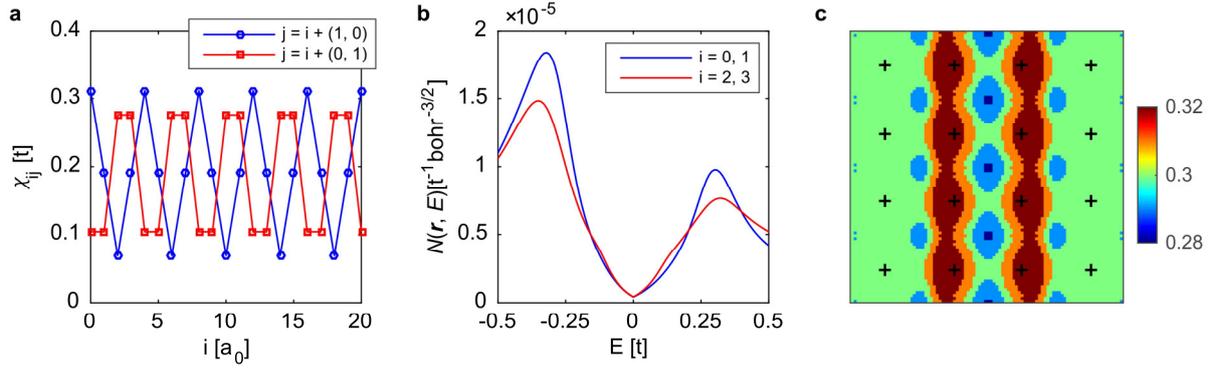

**Supplementary Figure 8. Characteristics of a *d*-form factor bond density wave coexisting with uniform d-wave superconductivity (BDW+DSC).**
(**a**) NN bond fields $\chi_{ij}$, (**b**) continuum LDOS $N(\boldsymbol{r}, E)$ at two inequivalent Cu sites, and (**c**) gap map $\Delta_1(\boldsymbol{r})$ (in units of $t$) over $4a_0 \times 4a_0$ area in BDW+DSC state. The BDW+DSC state is constructed "by hand" using a model Hamiltonian which consists of the normal state term and uniform DSC term derived from the uniform DSC state solution of the RMFT t-J model at $T = 0.01t$, and *d*-form factor BDW term with the same amplitude as the DSC pair field, see the Supplementary Information Section D of Ref. [11].

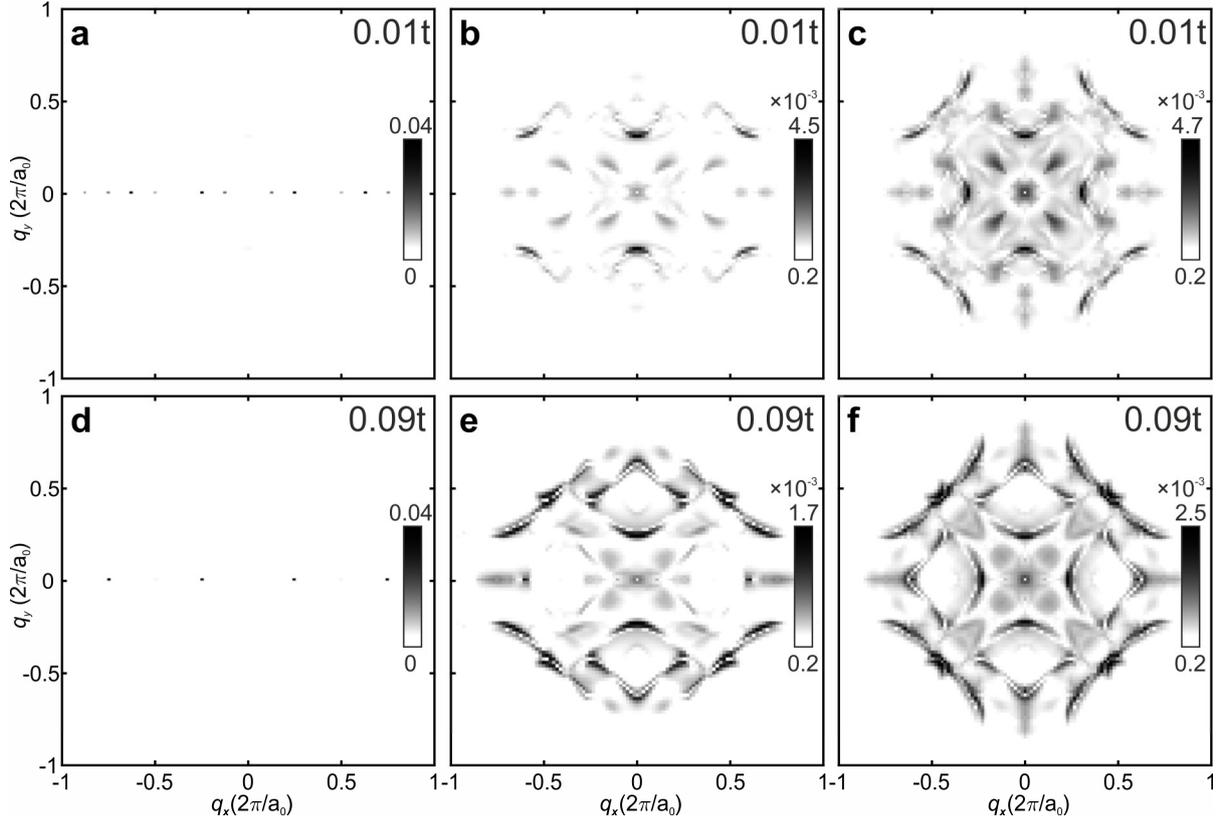

**Supplementary Figure 9.**
**a**. BQPI $Z(\boldsymbol{q}, E)$-map at energy $E = 0.03t$ in PDW+DSC state at temperature $T = 0.01t$ obtained using parameters same as in Fig. 4 of the main-text. Largest intensity occurs at non-dispersing charge order Bragg-peaks $\boldsymbol{q} = \pm n\boldsymbol{Q}_P, n = 0, 1, 2, \dots, 7$.
**b**. Same as in (**a**) with charge order Bragg-peaks suppressed for a better visualization of QPI wavevectors emerging from impurity scattering and to account for short-range discommensurate nature of charge order seen in the experiments.
**c**. Symmetrized map obtained by adding the map in (**b**) and its 90° rotated version.
**d**. BQPI $Z(\boldsymbol{q}, E)$-map at energy $E = 0.03t$ in pure PDW state at temperature $T = 0.09t$. Largest intensity occurs at non-dispersing charge order Bragg peaks $\boldsymbol{q} = \pm n(2\boldsymbol{Q}_P), n = 0, 1, 2, 3$.
**e**. Same as in (**d**) with charge order Bragg-peaks suppressed.
**f**. Symmetrized map obtained by adding the map in (**e**) and its 90° rotated version.

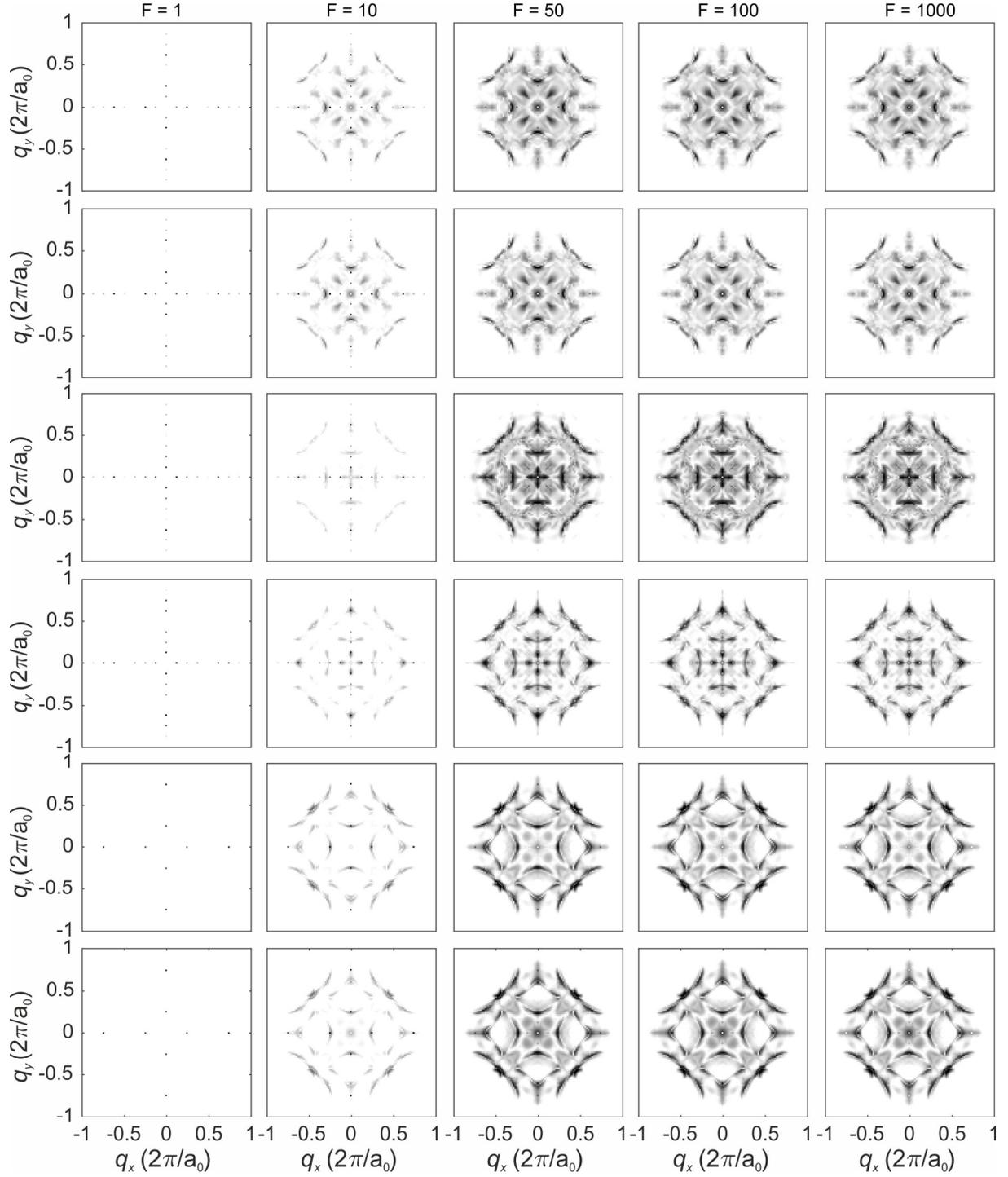

**Supplementary Figure 10.** Energy-integrated BQPI $\Lambda_P(\boldsymbol{q}, \Delta_0)$ in PDW+DSC state at $T = 0.01t, 0.02t, 0.04t, 0.05t$, and in pure PDW state at $T = 0.085t, 0.09t$, for various values of charge order Bragg peak suppression factors $F$, mentioned on the top of each column.

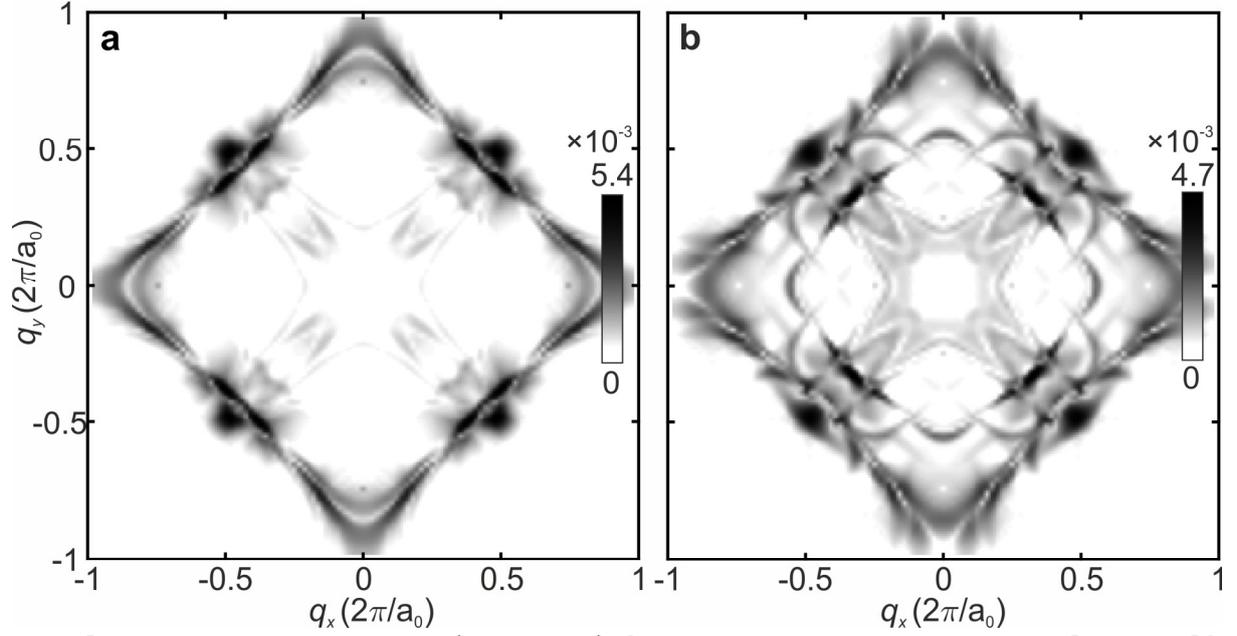

**Supplementary Figure 11.** $\Lambda_C(\boldsymbol{q}, 20\,\text{meV})$ **for** $4\boldsymbol{a_0}$ **CDW state constructed non-self-consistently at temperature** $T = 0.09t$.

**a.** $\Lambda_C(\boldsymbol{q}, 20\,\text{meV})$ for a CDW state constructed by setting the pair field to zero in the pure PDW state that is obtained self-consistently at $T = 0.09t$.

**b.** $\Lambda_C(\boldsymbol{q}, 20\,\text{meV})$ for a CDW state constructed by taking the normal state Hamiltonian from the uniform DSC state solution at $T = 0.09t$ and, subsequently, adding a $d$-form factor charge density wave term.

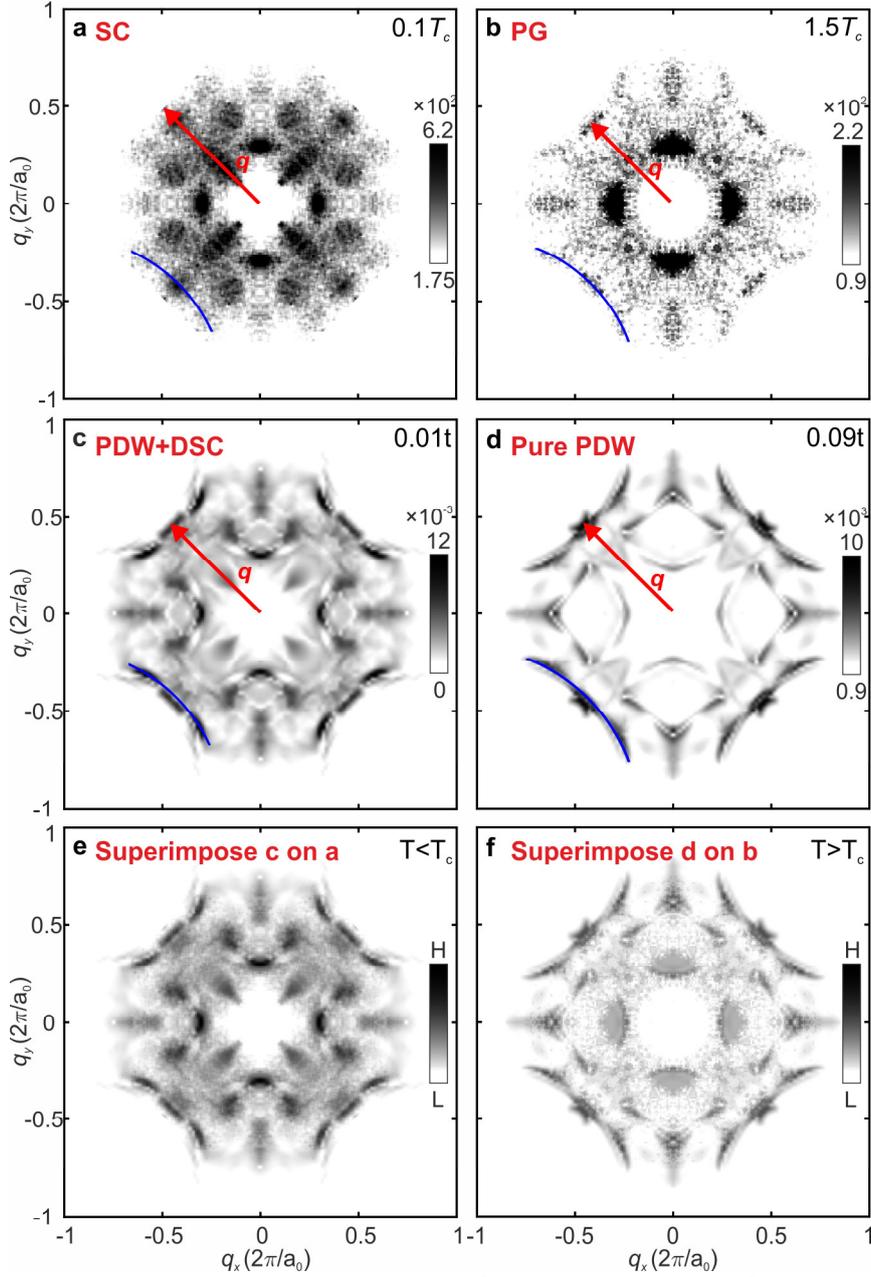

**Supplementary Figure 12. Discrimination of QPI signature of the pseudogap phase from the superconducting state**.

**a**. Measured $\Lambda(\boldsymbol{q}, 20\,\text{meV})$ in the superconducting state at $T = 0.1T_c$. The contrast of the QPI pattern is maximized for better comparison between the superconducting state and the pseudogap phase. The arc is highlighted by a blue curve. The lobe extending in the nodal direction is indicated by a red arrow.

**b**. Measured $\Lambda(\boldsymbol{q}, 20\,\text{meV})$ in the pseudogap phase at $T = 1.5T_c$.

**c**. Predicted $\Lambda_P(\boldsymbol{q}, 20\,\text{meV})$ of the PDW +DSC state at $T = 0.01t$. The lobe extending in the nodal direction is a signature of the presence of DSC component.

**d**. Predicted $\Lambda_P(\boldsymbol{q}, 20\,\text{meV})$ of the pure PDW state at $T = 0.09t$. The lobe in the nodal direction disappears above $T_c$, as a consequence of vanishing DSC component.

**e**-**f**. Superimposition of the calculated $\Lambda_P(\boldsymbol{q}, 20\,\text{meV})$ onto the measured $\Lambda(\boldsymbol{q}, 20\,\text{meV})$ shows excellent coincidence of the positions of dominant QPI features.

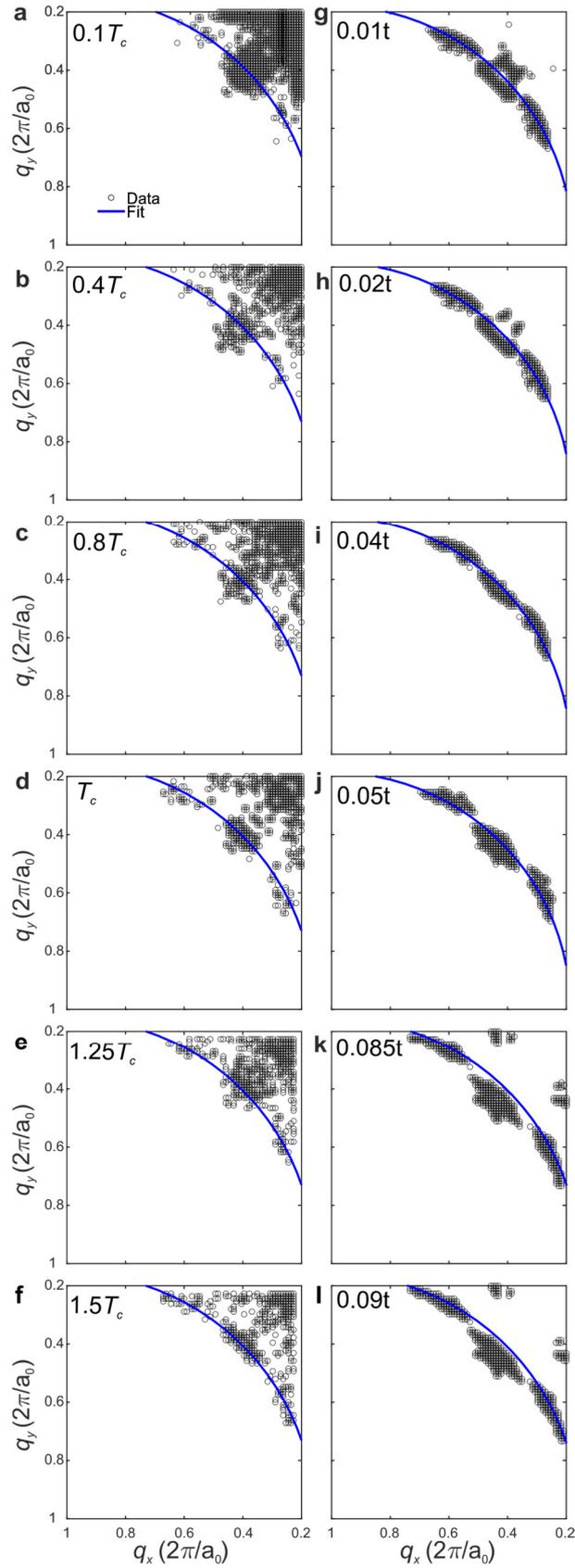

**Supplementary Figure 13.** The arc feature in $\Lambda(\boldsymbol{q}, 20 \text{ meV})$ is fit by a circle about $(\pm 1, \pm 1) \ 2\pi/a_0$ point. The angle subtended by this arc is measured versus temperature.

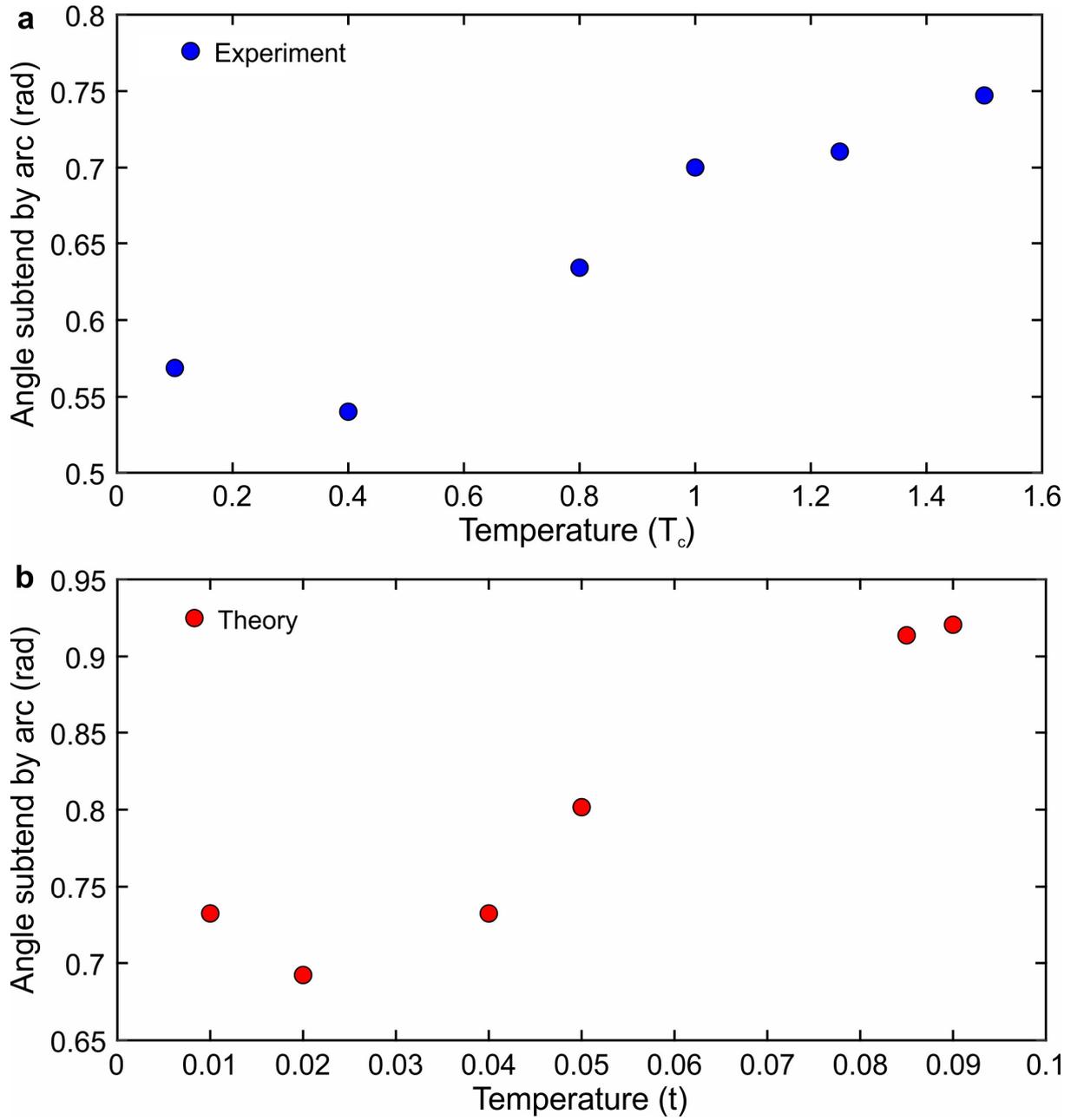

**Supplementary Figure 14.** The temperature dependence of the arc feature in (**a**) measured $\Lambda(\boldsymbol{q}, 20 \text{ meV})$ and (**b**) predicted $\Lambda_P(\boldsymbol{q}, 20 \text{ meV})$. The arc extension grows with temperature through $T_c$ in both measurement and predictions.

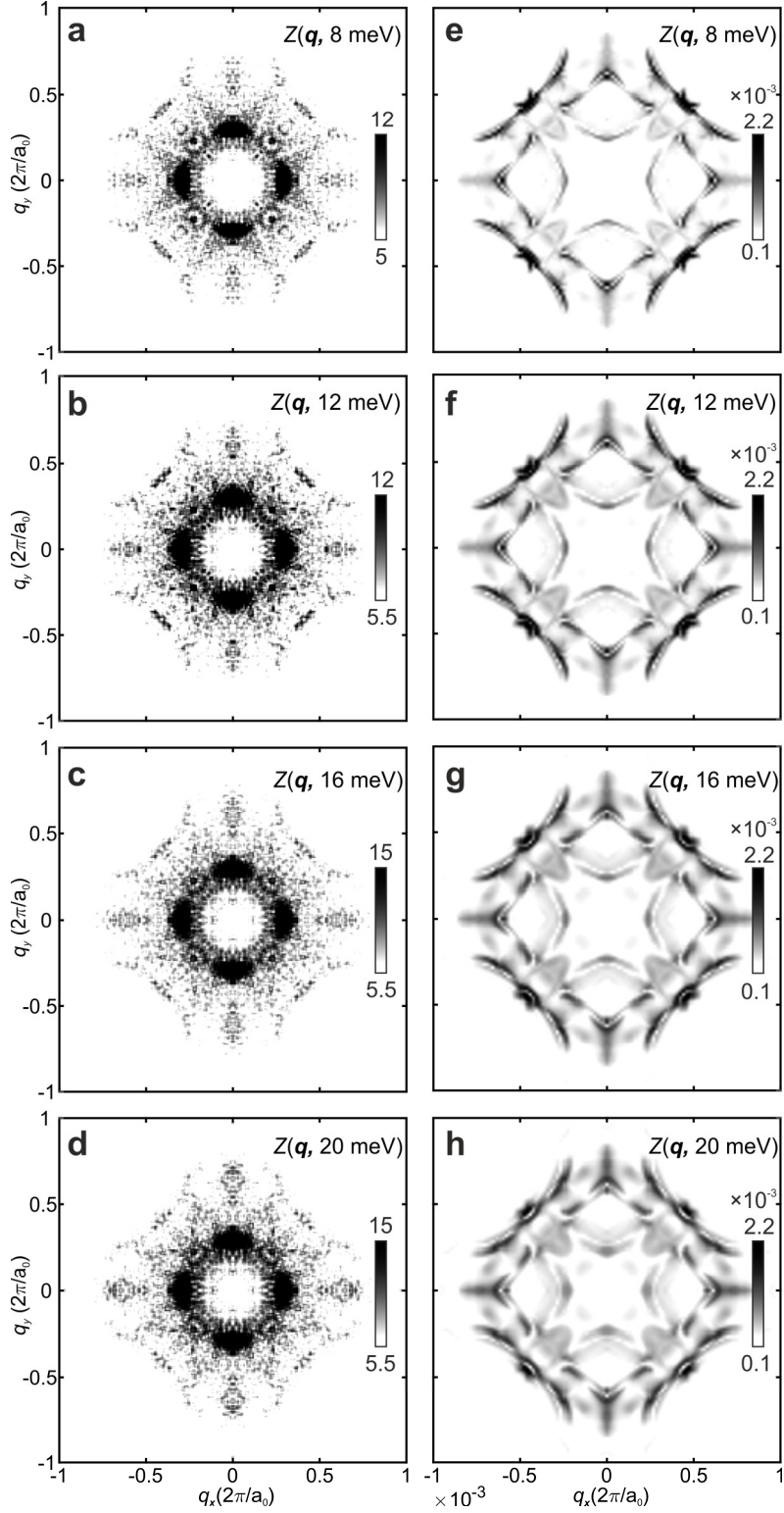

**Supplementary Figure 15. Energy dependence of quasiparticle inference $Z(\boldsymbol{q}, V)$ in the pseudogap phase and the pure PDW state**.
**a-d**. The experimental $Z(\boldsymbol{q}, V)$ maps measured at 55 K.
**e-h**. The theoretical $Z(\boldsymbol{q}, V)$ maps predicted for the pure PDW state $T = 0.09t$.